\documentclass[11pt,a4paper]{article}
\usepackage[T1]{fontenc}
\usepackage{amsfonts,amsmath,amssymb,amsthm}
\usepackage{sansmath}
\usepackage{graphicx}
\usepackage{hyperref}
\usepackage[a4paper,includeheadfoot,margin=2.54cm]{geometry}
\usepackage[affil-it]{authblk}

\newcommand{\heading}[1]{\medskip\par\noindent{\bf #1}}

\newcommand*{\proofclaimtext}{Proof}
\newenvironment{proofclaim}[1][\proofclaimtext]{\begin{proof}[#1]}{\end{proof}}

 \def\calB{{\cal B}} \def\calC{{\cal C}} 
  \def\calG{{\cal G}} \def\calH{{\cal H}}
   
  \def\calO{{\cal O}} \def\calP{{\cal P}}
 \def\calR{{\cal R}} \def\calS{{\cal S}} 
   \def\calX{{\cal X}}

\def\cNP{\hbox{\rm \sffamily NP}}
\def\cGI{\hbox{\rm \sffamily GI}}
\def\cW#1{\hbox{\rm \sffamily W[#1]}}
\def\cXP{\hbox{\rm \sffamily XP}}
\def\cFPT{\hbox{\rm \sffamily FPT}}
\def\cAPX{\hbox{\rm \sffamily APX}}

\def\hid{{\textsc{IntDim(1,3)}}}
\def\PrColExt#1#2{{\textsc{PrColExt($#1,#2$)}}}

\def\int{\hbox{\rm \sffamily INT}}

\def\ca{\hbox{\rm \sffamily CARC}}

\def\chor{\hbox{\rm \sffamily CHOR}}

\def\split{\hbox{\rm \sffamily SPLIT}}

\def\graphs#1{\hbox{\sffamily $#1$-GRAPH}}

\def\cgraphs{\hbox{\rm \sffamily CACTUS-GRAPH}}

\DeclareMathOperator{\tw}{tw}

\newtheorem{theorem}{Theorem}
\newtheorem{lemma}{Lemma}[section]

\newtheorem{corollary}[theorem]{Corollary}

\newtheorem{claim}{Claim}[section]

\theoremstyle{definition}

\newtheorem{problem}{Problem}

\title{
On $H$-topological intersection graphs
\thanks{
This paper is the combination and extension of the conference versions which appeared at
WG~2017~\cite{ChaplickTVZ16} and Eurocomb~2017~\cite{ChaplickZ17}.}}

\author[1]{Steven Chaplick}
\author[2]{Martin T\"{o}pfer}
\author[2]{Jan Voborn\'{i}k}
\author[2]{Peter Zeman\thanks{Supported by GAUK~1224120 and by GA\v{C}R 19-17314J.}}
\affil[1]{Department of Data Science and Knowledge Engineering, 
    Maastricht University, The Netherlands, \texttt{s.chaplick@maastrichtuniversity.nl}.}
\affil[2]{Department of Applied Mathematics, Faculty of Mathematics and Physics, Charles University, Czech Republic, \texttt{\{topfer,vobornik,zeman\}@kam.mff.cuni.cz}.}
\date{}

\begin{document}

\maketitle

\begin{abstract}
Bir\'{o}, Hujter, and Tuza (1992) introduced the concept of $H$-graphs, intersection graphs of
connected subgraphs of a subdivision of a graph $H$.
They are
related to and generalize many important classes of geometric intersection graphs, e.g., interval
graphs, circular-arc graphs, split graphs, and chordal graphs.  Our paper starts a new line of
research in the area of geometric intersection graphs by studying several classical computational problems on $H$-graphs: recognition, graph
isomorphism, dominating set, clique, and colorability.

We negatively answer the 25-year-old question of Bir\'{o}, Hujter, and Tuza which asks
whether $H$-graphs can be recognized in polynomial time, for a fixed graph $H$. We prove that it is
$\cNP$-complete if $H$ contains the \emph{diamond graph} as a minor. On the positive side, we
provide a polynomial-time algorithm recognizing $T$-graphs, for each fixed tree $T$. For the special
case when $T$ is a star $S_d$ of degree $d$, we have an $\calO(n^{3.5})$-time algorithm.

We give $\cFPT$- and $\cXP$-time algorithms solving the minimum dominating set problem on
$S_d$-graphs and $H$-graphs, parametrized by $d$ and the size of $H$, respectively.
The algorithm for $H$-graphs adapts to an $\cXP$-time algorithm for the independent set and the
independent dominating set problems on $H$-graphs.

If $H$ contains the \emph{double-triangle} as a minor, we prove that the graph isomorphism problem
is \cGI-complete and that the clique problem is \cAPX-hard. On the positive side, we show that the
clique problem can be solved in polynomial time if $H$ is a cactus graph. Also, when a graph has
a Helly $H$-representation, the clique problem is polynomial-time solvable.

Further, we show that both the $k$-clique and the list $k$-coloring problems are solvable in
\cFPT-time on $H$-graphs, parameterized by $k$ and the treewidth of $H$. In fact, these results
apply to classes of graphs with treewidth bounded by a function of the clique number. 

We observe that $H$-graphs have at most $n^{O(\|H\|)}$ minimal separators which allows us
to apply the meta-algorithmic framework of Fomin, Todinca, and Villanger (2015) to show that for
each fixed $t$, finding a \emph{maximum induced subgraph of treewidth $t$} can be done in polynomial
time. In the case when $H$ is a cactus, we improve the bound to $O(\|H\|n^2)$. 
\end{abstract}

\newpage

\tableofcontents

\newpage

\section{Introduction}

An intersection representation $\calR$ of a graph $G$ is a collection of sets $\{R_v : v \in V(G)\}$
such that $R_u \cap R_v \neq \emptyset$ if and only if $uv \in E(G)$.  Many important classes of
graphs arise from restricting the sets $R_v$ to geometric objects (e.g., intervals, circular-arcs,
convex sets, planar curves).  The study of these geometric representations has been motivated
through various application domains. For example, intersection graphs of planar curves relate to
circuit layout problems~\cite{sinden1966,BradyS1990}, interval graphs relate to scheduling
problems~\cite{roberts1978graph} and can be used to model biological problems~(see, e.g.,
\cite{JosephMT92}), and intersection representations of convex sets relate to the study of wireless
networks~\cite{HusonS1995-wireless}.

We study \emph{$H$-graphs}, intersection graphs of connected subgraphs of a subdivision of a fixed graph $H$, introduced by Bir\'{o}, Hujter, and
Tuza~\cite{biro1992precoloring}. We answer their open question concerning the problem of recognition
of $H$-graphs and further start a new line of research in the area of geometric intersection graphs,
by studying $H$-graphs from the point of view of fundamental computational problems of theoretical
computer science: recognition, graph isomorphism, dominating set, clique, and colorability. We
begin by discussing several closely related graph classes.

\begin{figure}[b]
\centering
\includegraphics{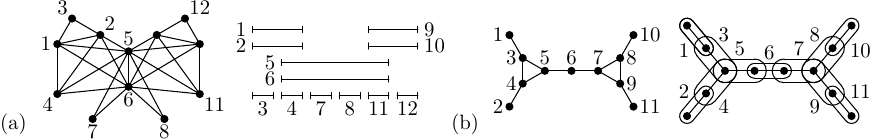}
\caption{
(a) An interval graph and one of its interval representations. (b) A chordal graph and one of its
representations as an intersection graph of subtrees of a tree.
}
\label{fig:int_chor_ex}
\end{figure}

\emph{Interval graphs} ($\int$) form one of the most studied and well-understood classes of
intersection graphs. In an \emph{interval representation}, each set $R_v$ is a closed interval of
the real line; see Fig.~\ref{fig:int_chor_ex}a.  A primary motivation for studying interval graphs
(and related classes) is the fact that many important computational problems can be solved in linear
time on them; see for example~\cite{PQ_trees,chang1998efficient,lueker1979linear}.

\emph{Chordal graphs} ($\chor$) were originally defined as the graphs without induced cycles of length greater than three.
Equivalently, as shown by Gavril~\cite{gavril1974intersection}, a graph is chordal if and only if it
can be represented as an intersection graph of subtrees of some tree; see
Fig.~\ref{fig:int_chor_ex}b. This immediately implies that $\int$ is a subclass of the chordal
graphs.

The recognition problem can be solved in linear time for $\chor$~\cite{rose1976algorithmic},
and such algorithms can be used to generate an intersection representation by subtrees of a tree.
However, asking for special host trees can be more difficult. For example, when the desired tree $T$
is a part of the input, deciding whether $G$ is a $T$-graph is \cNP-complete~\cite{DBLP:journals/tcs/KlavikKOS15}.
Additionally, some other important computational problems, for example the dominating
set~\cite{booth1982dominating} and graph isomorphism~\cite{lueker1979linear}, are harder on chordal
graphs than on interval graphs.

One can ask related questions about having ``nice'' tree representations of a given chordal graph.
For example, for a given graph $G$, if one would like to find a tree $T$ with the fewest leaves such
that $G$ is a $T$-graph, it can be done in polynomial time~\cite{habib2012reduced}, this is known as
the \emph{leafage} problem.  However, for any fixed $d \geq 3$, if one would like to find a tree
$T$ where $G$ is a $T$-graph and, for each vertex $v$, the subtree representing $v$ has at most $d$
leaves, the problem again becomes \cNP-complete~\cite{chaplick2014vertex}, this is known as the
\emph{$d$-vertex leafage} problem.  The minimum vertex leafage problem can be solved in
$n^{O(\ell)}$-time via a somewhat elaborate enumeration of \emph{minimal}\footnote{where each node
of $T$ corresponds to a maximal clique of $G$} tree representations of $G$ with exactly $\ell$
leaves where $\ell$ is the leafage of $G$~\cite{chaplick2014vertex}.

\emph{Split graphs} (\split) form an important subclass of chordal graphs.  These are the graphs
that can be partitioned into a clique and an independent set. Note that every split graph can be
represented as an intersection graph of subtrees of a \emph{star} $S_d$, where $S_d$ is the complete
bipartite graph $K_{1,d}$.

\emph{Circular-arc graphs} ($\ca$) naturally generalize interval graphs.  Here, each set
$R_v$ corresponds to an arc of a circle. The \emph{Helly circular-arc graphs} form an important 
subclass of circular-arc graphs.  A graph $G$ is a \emph{Helly circular-arc graph} if the
collection of circular-arcs $\calR = \{R_v\}_{v \in V(G)}$ satisfies the \emph{Helly property},
i.e., in each sub-collection of $\calR$ whose sets pairwise intersect, the common intersection is
non-empty. Interestingly, it is \cNP-hard to compute a minimum coloring for Helly circular-arc
graphs~\cite{HellyCircArcColoring-1996}.

\subsection{\texorpdfstring{$H$}{H}-graphs}
\label{sec:h-graphs}

Bir\'{o}, Hujter, and Tuza~\cite{biro1992precoloring} introduced \emph{$H$-graphs}. Let $H$ be a
fixed graph. A graph $G$ is an \emph{intersection graph of $H$} if it is an intersection graph of
connected subgraphs of $H$, i.e., the assigned subgraphs $H_v$ and $H_u$ of $H$ share a vertex if
and only if $uv \in E(G)$.

A \emph{subdivision} $H'$ of a graph $H$ is obtained when the edges of $H$ are replaced by
internally disjoint paths of arbitrary lengths. A graph $G$ is a \emph{topological intersection
graph} of $H$ if $G$ is an intersection graph of a subdivision $H'$ of $H$. We say that $G$ is an
\emph{$H$-graph} and the collection $\{H'_v : v \in V(G)\}$ of connected subgraphs of $H'$ is an
\emph{$H$-representation} of $G$. The class of all $H$-graphs is denoted by $\graphs{H}$.
Alternatively, we can view $H$-graphs geometrically as intersection graphs of connected subregions
of a one-dimensional simplicial complex (this is a topological definition of a graph).  We have the
following relations:
$$\int = \graphs{K_2},\quad \ca = \graphs{K_3},$$
$$\split \subsetneq \bigcup_{d =
2}^{\infty}\graphs{S_d},\quad \chor = \bigcup_{\text{Tree}\ T}\graphs{T}.$$

\heading{Motivation.}
It is easy to see that every graph $G$ is an $H$-graph for an appropriate choice of $H$ (e.g., by
taking $H = G$). In this sense, the families of $H$-graphs provide a parameterized view through
which we can study all graphs.  We also mentioned that several important computational problems are
polynomial on interval (the most basic class of $H$-graphs), but are hard on chordal graphs.  This
inspires the question of when we can use this parameterization to provide a refined understanding of
computational problems.  Of course, to approach this problem, we first need to observe some
relations among the classes of $H$-graphs and related well-studied graph classes. 

For any pair of (multi-)graphs $H_1$ and $H_2$, if $H_1$ is a \emph{minor} of $H_2$, then
$\graphs{H_1} \subseteq \graphs{H_2}$. Moreover, if $H_1$ is a subdivision of $H_2$, then
$\graphs{H_1} = \graphs{H_2}$.  Specifically, we have an infinite hierarchy of graph classes between
interval and chordal graphs since for every tree $T$ with at least one edge, $\int \subseteq \graphs{T} \subsetneq \chor$.
This motivates the study of the above mentioned problems on $T$-graphs, for a fixed tree~$T$.

We note a dichotomy regarding computing a minimum coloring on \graphs{H}.  Namely, if $H$ contains a
cycle, then computing a minimum coloring on $\graphs{H}$ is already \cNP-hard even for the subclass
of Helly $H$-graphs~\cite{HellyCircArcColoring-1996}. On the other hand, when $H$ is acyclic, a
minimum coloring can be computed in linear time since $\graphs{H}$ is a subclass of $\chor$.

Bir\'{o}, Hujter, and Tuza originally introduced $H$-graphs in the context of the \emph{$(p,k)$
pre-coloring extension problem} (\PrColExt{p}{k}). In this problem, the input is a graph $G$
together with a $p$-coloring of $W \subseteq V(G)$, and the goal is to find a proper $k$-coloring of
$G$ extending this \emph{pre-coloring}.  Bir\'{o}, Hujter, and Tuza~\cite{biro1992precoloring}
provide an $\cXP$ (in $k$ and $\|H\|$) algorithm to solve \PrColExt{k}{k} on $H$-graphs.  Bir\'{o},
Hujter, and Tuza asked the following question which we answer negatively.

\begin{quote}[Bir\'{o}, Hujter, and Tuza~\cite{biro1992precoloring}, 1992]
\label{prob:tuza}
Let $H$ be an arbitrary fixed graph. Is there a polynomial algorithm testing whether a given graph
$G$ is an $H$-graph?
\end{quote}

\subsection{Our results}
\label{sec:results}

We give a comprehensive study of $H$-graphs from the point of view of several important problems of
theoretical computer science: recognition, graph isomorphism, dominating set, clique, and
colorability. We focus on five collections of classes of graphs. In particular, $\graphs{S_d}$,
$\graphs{T}$, $\graphs{C}$, Helly $\graphs{H}$, and $\graphs{H}$, where $S_d$ is the star of degree
$d$, $T$ is a tree, $C$ is a cactus, and $H$ is an arbitrary graph. Our results are displayed in
Table~\ref{fig:tabulka}. The following list provides a summary of our results and should help the
reader to navigate through the paper:

\begin{itemize}
\item
\emph{Recognition.}
In Section~\ref{sec:hardness} we negatively answer the question of Bir\'{o}, Hujter, and Tuza.  We
prove that recognizing $H$-graphs is $\cNP$-complete if $H$ is not a cactus
(Theorem~\ref{thm:diamond-hardness}). Equivalently this means that $H$ contains the diamond graph as
a minor. We do this by a reduction from the problem of testing whether the interval dimension of a
partial order of height $2$ is at most $3$. On the positive side, in Section~\ref{sec:recog-poly},
we give an $\calO(n^{3.5})$-time algorithm for recognizing $S_d$-graphs
(Theorem~\ref{thm:poly_star}), and we give a polynomial-time algorithm for recognizing $T$-graphs
(Theorem~\ref{thm:xp_tree}), for a fixed tree $T$.

\item
\emph{Dominating set.}
In Section~\ref{sec:domset}, we solve the problem of finding a minimum dominating set for
$S_d$-graphs in time  $\calO(dn (n+m)) + 2^{d}(d + 2^d)^{\calO(1)}$
(Theorem~\ref{thm:fpt_star_domination}) and for $H$-graphs in $n^{\calO(\|H\|)}$-time
(Theorem~\ref{thm:xp_domination}). The latter algorithm can be easily adapted to solve the maximum
independent set problem and minimum independent dominating set problem in $n^{\calO(\|H\|)}$-time
for $H$-graphs (Corollary~\ref{cor:mis_ids}).

\item
\emph{Clique.}
In Section~\ref{sec:clique}, we study the clique problem. We show that if $H$ contains the
\emph{double-triangle} $\Delta_2$ (see Fig.~\ref{fig:comp2sub}a) as a minor, then the clique problem
is \cAPX-hard for $H$-graphs (Theorem~\ref{thm:double-triangle->Subd2_inside}).  On the positive
side, we solve the clique problem in polynomial time for Helly $H$-graphs
(Theorem~\ref{thm:helly-h-clique}), and in the case when $H$ is a cactus
(Theorem~\ref{thm:clique-cactus}).

\item
\emph{Graph isomorphism.}
Theorem~\ref{thm:double-triangle->Subd2_inside} also gives that if $H$ contains the
\emph{double-triangle} $\Delta_2$ (see Fig.~\ref{fig:comp2sub}a) as a minor, then graph isomorphism
problem is $\cGI$-complete for $H$-graphs.

\item
\emph{$k$-coloring and $k$-clique.}
In Section~\ref{sec:treewidth}, we use treewidth based methods to provide an $\cFPT$-time algorithm
for finding a $k$-clique in an $H$-graphs (Theorem~\ref{thm:fpt-k-clique}) and an $\cFPT$-time
algorithm for $k$-coloring of $H$-graphs (Theorem~\ref{thm:fpt-k-list-col}).  In fact, these
results apply to more general graph classes formalized via the concept of a \emph{clique-treewidth property} 
(which is defined as in the \emph{parameter-treewidth properties} of \emph{bidimensionality theory}; 
see, e.g.,\cite{parametertreewidth2004}) and may be of independent interest. 

\item 
\emph{Minimal Separators.}
Finally, in Section~\ref{sec:min_sep}, we show that each $H$-graph has $n^{O(\|H\|)}$ minimal
separators (Theorem~\ref{thm:min_sep}) and, when $H$ is a cactus, we improve this bound to
$O(\|H\|n^2)$ (Theorem~\ref{thm:min_sep_cactus}). Thus, by the algorithmic framework of Fomin,
Todinca, and Villanger~\cite{FominTV15}, on $H$-graphs, we obtain a large class of problems
(including, e.g., feedback vertex set) which can be solved in \cXP-time (parameterized by $\|H\|$)
and polynomial time (in both $\|H\|$ and the size of the input graph) when $H$ is a cactus.
\end{itemize}

\begin{table}[t]
\centering
\includegraphics{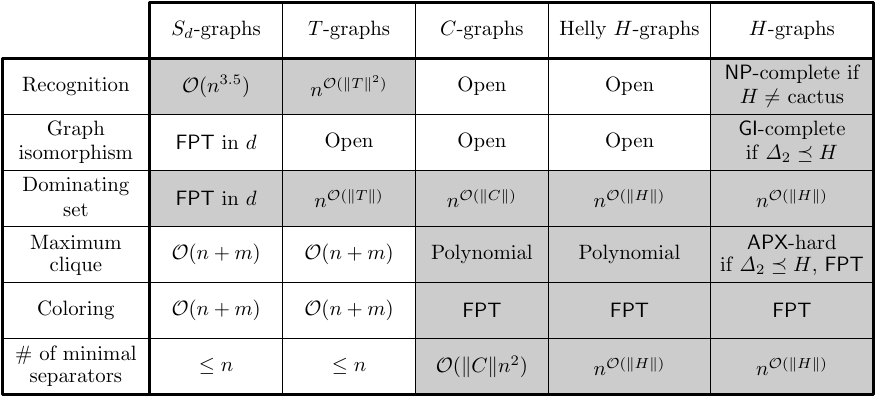}
\caption{
The table of the complexity of different problems for the four
considered classes. Our contributions are highlighted. Note: $A \preceq B$ denotes that $A$ is a
minor of $B$, and $\Delta_2$ denotes the double-triangle (see Fig.~\ref{fig:comp2sub}).
}
\label{fig:tabulka}
\end{table}

\heading{Open problems.}
Since all the sections are mostly self-contained, instead of including a separate section for open
problems and conclusions, we decided to include the open problems and possible future research
directions in the corresponding sections.

\heading{Recent developments.}
After the publication of the two conference articles~\cite{ChaplickTVZ16,ChaplickZ17} (which
this paper includes and extends), there have already been further developments regarding $H$-graphs~\cite{FominGR-2017arXiv,JaffkeKT-2017arXiv}.  The results contained in these articles
complement and build on our work regarding combinatorial optimization problems.  For instance, to
complement our \cXP-time algorithms for minimum dominating set and maximum independent set, Fomin,
Golovach, and Raymond~\cite{FominGR-2017arXiv} show that these problems are \cW{1}-hard, when
parameterized by $\|H\|$ and the desired solution size.  They additionally tighten our result
regarding the fixed parameter tractability of the $k$-Clique problem on $H$-graphs by showing that
this problem admits a polynomial size kernel in terms of both $\|H\|$ and the solution size.
Jaffke, Kwon, and Telle~\cite{JaffkeKT-2017arXiv} adapt the \cW{1}-hardness proof
from~\cite{FominGR-2017arXiv} for maximum independent set to additionally show that feedback vertex
set is also \cW{1}-hard. 
Only recently, the problem of testing isomorphism of $S_d$ graphs was solved in FPT time~\cite{AgaogluH20}.

\section{Preliminaries}

We assume that the reader is familiar with the following standard and parameterized computational
complexity classes: \cNP, \cXP, and \cFPT\ (see, e.g., \cite{cygan2015parameterized} for further
details). 

Let $G$ be an $H$-graph. 
For a subdivision $H'$ certifying $G \in \graphs{H}$, we use $H_v'$ to
denote the subgraph of $H'$ corresponding to $v \in V(G)$. 
The vertices of $H$ and $H'$ are called \emph{nodes}. 
By $\|H\|$ we denote the \emph{size} of $H$, i.e., $\|H\| = |V(H)| + |E(H)|$.

We refer to the degree~one nodes of $H$ as \emph{leaves} and the nodes degree at least three as \emph{branching points}. 
Note that, while we sometimes speak of degree two nodes in $H$, they are actually redundant since their presence or absence does not change $\graphs{H}$. 
As such, by thinking of $H$ as a multi-graph with loops one can nearly always avoid the need for any nodes of degree two (by contracting edges where one end point has degree two).
The exception here is the case of $H$ being a cycle which leads to the \emph{true } $H$ simply being a single vertex with one loop, i.e., this vertex has degree two. 
Of course, when $H$ is a tree, this works without the need for $H$ to be a multi-graph. 

We have some special notation for the case when $H$ is a tree. Let $a, b$ be two nodes of $H'$. 
By $P_{[a,b]}$ we denote the path from $a$ to $b$. 
Further, we define $P_{(a,b]} := P_{[a,b]} - a$, and $P_{[a,b)}$, $P_{(a,b)}$ analogously.

Let $S \subseteq G$. Then $G[S]$ is the subgraph of $G$ induced by $S$, and $G - S$ is the graph
obtained from $G$ by deleting the vertices in $S$ (together with the incident edges). For a graph
$G$, we assume $G$ has $n$ vertices and $m$ edges.

In 1965, Fulkerson and Gross proved the following fundamental characterization
of interval graphs by orderings of maximal cliques. It is used implicitly in
several proofs. 
\begin{lemma}[Fulkerson and Gross~\cite{maximal_cliques}] \label{lem:fulkerson_gross}
A graph $G$ is an interval graph if and only if there exists a linear ordering $\preceq$ of the
maximal cliques of $G$ such that for every $u \in V(G)$ the maximal cliques containing $u$ appear
consecutively in $\preceq$.
\end{lemma}

\heading{A remark on the size of subdivisions and membership in \cNP.}
As membership in $\graphs{H}$ is certified through the existence of an appropriate subdivision of
$H$, one might wonder just how large subdivision $H'$ is necessary to ensure that any
$n$-vertex $H$-graph $G$ has a representation by connected subgraphs of $H'$. Note that as long as the size of this subdivision is bounded by a polynomial in $n$, $H$-graph recognition does indeed belong to \cNP.  
We observe that it suffices to subdivide every edge of $H$ $2n$ times to
accommodate an $n$-vertex $H$-graph, i.e., without loss of generality the size of $H'$ is at most
$|V(H))| + 4n|E(H)|$.

To see this, we consider an edge $ab$ of $H$, and its corresponding path $a,c_1, \ldots, c_\ell, b$
in $H'$.  Observe that, for each vertex $v \in V(G)$, $H'_v$ has at most two leaves on this path.
Thus, if $\ell > 2n$, there must be a $c_i$ which does not contain any leaf of any $H'_v$. 
In particular, this $c_i$ can be contracted into its neighbour on the path while preserving the
representation of $G$. Therefore, it suffices to consider subdivisions of size $|V(H))| +
4n|E(H)|$ and, in particular, for every $H$, recognition of $H$-graphs is in $\cNP$.

\section{Recognition is hard if \texorpdfstring{$H$}{H} is not a cactus}
\label{sec:hardness}

In this section, we negatively answer a question posed by Bir\'{o}, Hujter, and
Tuza~\cite[Problem 6.3]{biro1992precoloring}.  Namely, we prove that testing whether a graph is an $H$-graph is
\cNP-complete when the \emph{diamond graph}\footnote{The diamond graph is obtained by
deleting an edge from a 4-vertex clique.} $D$ is a minor of $H$.
Note that this  sharply contrasts the polynomial time
solvability of the recognition problem for circular-arc graphs (i.e., when $H$ is a cycle). 
Before getting to the hardness proof itself, we first establish a technical (though rather straightforward to prove) lemma regarding the essentially unique (up to automorphism) $H$-representability of the \emph{3-subdivision} $H_3$ of $H$ as an $H$-graph. 
Namely, $H_3$ is obtained from $H$ by subdividing each edge exactly 3 times, that is, in $H_3$ we have one vertex $x_v$ for each vertex $v$ of $H$, and for each edge $e = uv$ of $H$ we have the path $x_u, x_{ue}, x_e, x_{ve}, x_v$. 

\begin{lemma}\label{lem:3-subdivision-blocker}
Let $H$ be any multi-graph without vertices of degree~2, and let $H_3$ be the \emph{3-subdivision} of $H$. 
The graph $H_3$ is an $H$-graph and, for every subdivision $H'$ certifying $H_3 \in \graphs{H}$ (via the representation $\{H'_x : x \in V(H_3)\}$, we have:
\begin{itemize}
\item For each non-leaf vertex $v$ of $H$, the representation $H'_{x_v}$ of the corresponding vertex $x_v$ in $H_3$ contains exactly one branching point $p$ of $H$ where the degree of $v$ (and $x_v$) and $p$ coincide.
\item For each edge $e = uv$ of $H$ and the corresponding path $x_u x_{ue}, x_e, x_{ve}, x_v$ in $H_3$, the representation $H'_{x_e}$ of $x_e$ is strictly contained within the subdivision of a single edge $zz'$ of $H$ such that for distinct edges $e,f$ of $H$ with corresponding ``middle'' vertices $x_e,x_f$ in $H_3$, $H'_{x_e}$ and $H'_{x_f}$ are contained within subdivisions of distinct edges of $H$. 
\end{itemize}
Moreover, each $H$-representation of $H_3$ defines an automorphism of $H$. 
\end{lemma}
\begin{proof}
We first note that this holds trivially for the case when $H$ is $K_1$ or $K_2$. 

We now observe that $H_3$ is indeed an $H$-graph. 
Let $H'$ be the 4-subdivision of $H$, that is, in $H'$ the edge $e = uv$ of $H$ becomes the path $y_u y_{ue}, z_{ue}, z_{ve}, y_{ve}, y_v$. 
For each vertex $v$ of $H$ with incident edges $\{e_1, \ldots, e_k\}$, we represent $x_v$ by the star $H'_v = H'[\{y_v, y_{ve_1}, y_{ve_2}, \ldots, y_{ve_k}\}]$. 
For each edge $uv$ of $H$, we represent:
\begin{itemize}
\item $x_{ue}$ by $H'_{x_{ue}} = $ the edge $y_{ue}z_{ue}$, 
\item $x_{e}$ by $H'_{x_e} = $ the edge $z_{ue}z_{ve}$, and
\item $x_{ve}$ by $H'_{x_{ve}} = $ the edge $z_{ve}y_{ve}$. 
\end{itemize} 
It is easy to see that this collection of subgraphs of $H'$ is indeed $H$-representation of $H_3$. 

So, we now consider an arbitrary $H$-representation $\{H'_{x} : x \in V(H_3)\}$ of $H_3$, where $H'$ is the subdivision of $H$ and establish the claimed properties. 

Suppose that there is a vertex $v$ of $H$ where $v$ has degree at least three (with incident edges $e_1, \ldots, e_{k}$) and $H'_{x_v}$ does not contain a branching point, i.e., all nodes in $H'_{x_v}$ have degree at most two. 
Now, since the neighborhood $\{x_{ve_1}, \ldots, x_{ve_k}$ of $x_v$ is an independent set (and $k \geq 3$), this implies that (without loss of generality), $H_{x_{ve_1}}$ is contained within $H'_{x_v}$. 
However, this now makes it impossible to represent $x_{e_1}$ since $H'_{x_{e_1}}$ should intersect $H'_{x_{ve_1}}$ but should not intersect $H'_{x_v}$. 
Thus, for each vertex $v$ of $H$ with degree at least three, $H'_{x_v}$ contains a branching point. 
Note that no branching point can occur in two such $H'_{x_u}$ and $H'_{x_v}$, thus, the vertices of degree at least three are bijectively mapped to the branching points. 
Finally, since we now know that $H'_{x_v}$ contains exactly one branching point, we remark that the degree of this branching point must match be at least the degree of $x_v$ as otherwise some $H'_{x_{ve_i}}$ would be contained in $H'_{x_v}$ contradicting the $H$-representation at hand. 
Thus, indeed the degree of this branching point must coincide with the degree of $x_v$ (and $v$). 

Now consider any edge $e = uv$ of $H$. Observe that $H'_{x_e}$ cannot contain any branching points since each branching point is contained in a representation $H'_{x_v}$ where $x_v$ is not a neighbor of $x_e$. 
Thus, $H'_{x_e}$ is indeed contained within the subdivision of an edge $pq$ of $H$, and in particular when $p$ ($q$) is a branching point, then without loss of generality $H'_{x_u}$ ($H'_{x_v}$) contains $u$ ($p$). 
Observe that, when $u$ has degree at least three, $H'_{x_u} \cup H'_{x_{ue}} \cup H'_{x_{e}}$ consists of a subpath of the subdivision of $pq$ in $H'$ that includes $p$ and as such, for any edge $f$ distinct from $e$, $H'_{x_f}$ must be contained within the subdivision of a different edge of $H$. 
Moreover, this means that for each edge $e = uv$ connecting two vertices of degree at least three, $x_e$ is indeed represented on the subdivision of an edge connecting the corresponding branching points. 
Also, when one of $u$ or $v$, say $v$ has degree one (i.e., $v$ is a leaf of $H$ and $u$ has degree at least three), then $H'_{x_e}$ is contained in the subdivision of an edge $pq$ incident to the branching point $p$ contained in $H'_{x_u}$ where $q$ has degree one in $H$. 
In particular, here we also have that $H'_{x_v}$ is contained in the subdivision of $pq$. 

Finally, based on these properties, we indeed have an automorphism of $H$ as required. 
\end{proof}

Our hardness proof stems from the \cNP-hardness of testing whether a partial order (poset) with \emph{height} one has \emph{interval dimension} at most three; shown by Yannakakis~\cite{yannakakis1982complexity}.  
We denote this problem by \hid.  
Note that having \emph{height one} means that every element of the poset is either minimal or maximal.

Consider a collection $I$ of closed intervals on the real line. 
A poset $\calP_I=(I,<)$ can be defined on $I$ by considering intervals $x,y \in I$ and setting $x < y$ if and only if the right endpoint of $x$ is strictly to the left of the left endpoint of $y$. 
A partial order $\calP$ is called an \emph{interval order} when there is an $I$ such that $\calP = \calP_I$. The \emph{interval dimension} of a poset $\calP=(P,<)$, is the minimum number of interval orders whose intersection is $\calP$, i.e., for elements $x,y\in P$, $x < y$ if and only if $x$ is before $y$ in all of the interval orders.  
Finally, the \emph{incomparability graph} $G_\calP$ of a poset $\calP=(P,<)$  is
the graph with $V(G) = P$ and $uv \in E(G_\calP)$  if and only if $u$ and $v$ are not comparable in $\calP$.

\begin{figure}[b!]
\centering
\includegraphics[scale=1]{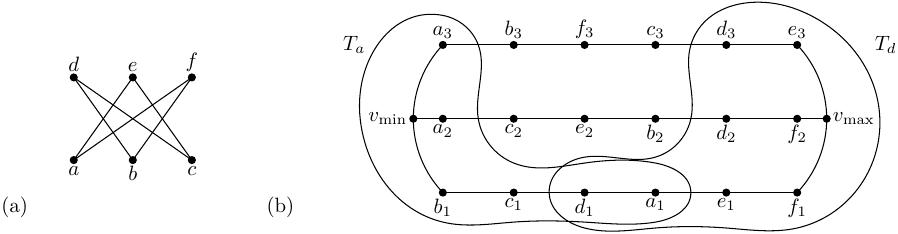}
\caption{
(a) A partially ordered set $\calP = (P,<)$ of height 1, interval dimension 3, but not 2. We define
the following interval orders:
$I_1 = l_{a} l_{b} l_{c} r_{b} r_{c} l_{d} r_{a} l_{e} l_{f} r_{d} r_{e} r_{f}$, 
$I_2 = l_{a} l_{b} l_{c} r_{a} r_{c} l_{e} r_{b} l_{d} l_{f} r_{d} r_{e} r_{f}$, and 
$I_3 =l_{a} l_{b} l_{c} r_{a} r_{b} l_{f} r_{c} l_{d} l_{e} r_{d} r_{e} r_{f}$,
where $[l_a,r_a]$ represents an interval corresponding to $a \in P$.  Note that $\calP_{I_1} \cap
\calP_{I_2} \cap \calP_{I_3} = \calP$.
(b) An illustration of part of the $D$-representation. Here, $T_a$ and $T_b$ indicate the subgraphs representing the elements $a$ and $b$. 
}
\label{fig:diamond}
\end{figure}

Note that if $\calP$ has height one, then $G_\calP$ is the \emph{complement of a bipartite graph}. 
The vertices $V(G_\calP)$ naturally partition into two cliques $K_{\max}$ and $K_{\min}$,
containing the maximal and the minimal elements of $\calP$, respectively. 
An example depicting a $D$-representation of a specific $\calP$ is provided in Fig.~\ref{fig:diamond}, where $D$ is the diamond graph.
With these definitions and the prior lemma in place, we now prove the theorem of this section. 

\begin{theorem}
\label{thm:diamond-hardness}
Testing if $G$ is an $H$-graph is \cNP-complete if the diamond graph $D$ is a minor of H.
\end{theorem}

\begin{proof}
The proof is split into two parts.
In the Part~1, we prove the essential case which shows that testing whether $G$ is an $D$-graph is \cNP-hard.
This argument is generalized in Part~2 to the case when $H$ contains $D$ as a minor. 

\heading{Part 1: $H$ is the diamond.}
First, we summarize the idea behind our proof. 
As stated above, we will encode an instance $\calP$ of \hid\ as an instance of membership testing in \graphs{D}.  
For a given height 1 poset $\calP$, we construct its incomparability graph $G_\calP$, slightly augment $G_\calP$ to get a graph $G$, and show that $G$ is in \graphs{D} if and only if the interval dimension of $\calP$ is at most 3. 
In particular, a ``middle'' part of the three paths connecting the two degree~3 vertices in $D$ will encode the three interval orders whose intersection is $\calP$. 

Note that, we consider $H$ as the multi-graph consisting of three parallel edges $e_a, e_b, e_c$ between two nodes $v_{\min}$ and $v_{\max}$
To construct $G$, we use the graph $H_3$ which the 3-subdivision of $H$. 
Namely, $H_3$ has two vertices $u_{\min}$ and $u_{\max}$ of degree three and nine vertices $a_1, a_2, a_3,$  $b_1, b_2, b_3,$ $c_1, c_2, c_3$ of degree two where $u_{\min}, \aleph_1, \aleph_2, \aleph_3, u_{\max}$ is a path for each $\aleph \in \{a,b,c\}$. 
Note that, by Lemma~\ref{lem:3-subdivision-blocker}, without loss of generality, $H_3$ is an $H$-graph where in every $H$-representation of $H_3$, say on a subdivision $H'$ of $H$, we have: 
\begin{itemize}
\item $H'_{u_{\min}}$ contains $v_{\min}$ and $H'_{u_{\max}}$ contains $v_{\max}$, 
\item For each $\aleph \in \{a,b,c\}$, $H'_{\aleph_2}$ is contained in the subdivision of $e_\aleph$. 
\end{itemize}
It is within these $H'_{\aleph_2}$ paths that we will see the interval orders. 

We are now ready to construct our graph $G$ from $H_3$ and the graph $G_\calP$ of a given height one poset $\calP=(P,<)$, recall that $K_{\min}$ and $K_{\max}$ denote cliques on the minima and maxima  of $\calP$ respectively. 
Let $V_{\min} = \{u_{\min}, a_1, a_2, b_1, b_2, c_1, c_2\}$ and let $V_{\max} = \{u_{\max}, a_3, a_2, b_3, b_2, c_3, c_2\}$. 
The graph $G$ is the union of $G_\calP$ and $H_3$ where, additionally, each vertex of $K_{\min}$ is adjacent to each vertex of $V_{\min}$ and each vertex of $K_{\max}$ is adjacent to each vertex of $V_{\max}$. 

\begin{claim}
$\calP$ has interval dimension at most $3$ if and only if $G$ is a $H$-graph. 
\end{claim}

\begin{proofclaim}
For the reverse direction, consider an $H$-representation of $G$ on a subdivision $H'$ of $H$.  
As remarked above, by Lemma~\ref{lem:3-subdivision-blocker}, $H'_{u_{\min}}$ contains the node $v_{\min}$ and $H'_{u_{\max}}$ contains the node $v_{\max}$. 
The minimal elements of $\calP$ are not adjacent to the vertices of $u_{\max}$. 
Therefore, for each $x \in K_{\min}$, $H'_{x}$ cannot contain $v_{\max}$, i.e., $H'_x$ is a subtree of $H' - \{v_{\max}\}$. 
In particular, for each of the three $(v_{\min},v_{\max})$ paths $A,B,C$ in $H'$, $H'_{x}$ defines one (possibly empty) subpath/interval (originating in $v_{\min}$). 
Similarly, for each $y \in K_{\max}$,  $H'_y$ cannot contain $v_{\min}$ and as such $H'_{y}$ defines, for each of $A,B,C$, one subpath (originating in $v_{\max}$). 
It is easy to see that these intervals provide the interval orders $\calP_{I_A}$, $\calP_{I_B}$, and $\calP_{I_C}$ such that $\calP_{I_A} \cap
\calP_{I_B} \cap \calP_{I_C} = \calP$.

For the forward direction, let $I_1$, $I_2$, $I_3$ be sets of intervals such that $\calP = \calP_{I_1} \cap \calP_{I_2} \cap \calP_{I_3}$.  
We assume that each interval in $I_i$ is labelled according to the corresponding element of $\calP$.  
Further, we assume that the intervals corresponding to the minimal elements have their left endpoints at $0$ and their right endpoints are integers
in the range $[0,n-1]$. 
Similarly, we assume that the intervals corresponding to the maximal elements have their right endpoints at $n$ and their left endpoints are integers in the range $[1,n]$. 
With this in mind, for each minimal element $x$ and each $i \in \{1,2,3\}$, we use $x_i$ to denote the right endpoint of its interval in $I_i$, and for each maximal element $y$ and each $i \in \{1,2,3\}$, we use $y_i$ to denote the left endpoint of its interval in $I_i$. 

Let $H'$ be the subdivision of $H$ obtained by subdividing the three $v_{\min}v_{\max}$ edges
$n+5$ times. 
We label the three $(v_{\min},v_{\max})$-paths in $H'$ as follows: 
\begin{itemize}
\item $v_{\min}, \alpha_{\min}, \alpha'_{\min}, \alpha_0, \alpha_1, \ldots, \alpha_n, \alpha'_{\max}, \alpha_{\max}, v_{\max}$,
\item $v_{\min}, \beta_{\min}, \beta'_{\min}, \beta_0, \beta_1, \ldots, \beta_n, \beta'_{\max}, \beta_{\max}, v_{\max}$, and 
\item $v_{\min}, \gamma_{\min}, \gamma'_{\min}, \gamma_0, \gamma_1, \ldots, \gamma_n, \gamma'_{\max}, \gamma_{\max}, v_{\max}$. 
\end{itemize}
We are now ready to describe an $H$-representation of $G$ on $H'$.
Each minimal element $x$ is represented by the minimal subtree of $H'$ which includes the nodes $v_{\min}, \alpha_{x_1}, \beta_{x_2},
\gamma_{x_3}$. 
Similarly, each maximal element $y$ is represented by the minimal subtree of $H'$
which includes the nodes $v_{\max}, \alpha_{y_1}, \beta_{y_2}, \gamma_{y_3}$. 
We can now see that the comparable elements of $\calP$ are represented by disjoint subgraphs of $H'$ and that the incomparable elements map to intersecting subgraphs. 
Finally, the vertices of $H_3$ are represented as follows:
\begin{itemize}
\item 
$u_{\min}$ is represented by the subtree induced by $v_{\min}$, $\alpha_{\min}$. 
$\beta_{\min}$, and $\gamma_{\min}$; 
analogously, $u_{\max}$ is represented by the subtree induced by $v_{\max}$, $\alpha_{\max}$,
$\beta_{\max}$, and $\gamma_{\max}$
\item
$a_1$, $b_1$, and $c_1$ are represented by the edges $\alpha_{\min}\alpha'_{\min}$, $\beta_{\min}\beta'_{\min}$, and
$\gamma_{\min}\gamma'_{\min}$, respectively;
analogously, $a_3$, $b_3$, and $c_3$ are represented by the edges $\alpha_{\max}\alpha'_{\max}$, $\beta_{\max}\beta'_{\max}$, and
$\gamma_{\max}\gamma'_{\max}$, respectively, and
\item
$a_2$ is represented by the path $\alpha'_{\min}, \alpha_0, \ldots, \alpha_n, \alpha'_{\max}$, and 
\item $b_2$ is represented by the path $\beta'_{\min}, \beta_0, \ldots, \beta_n, \beta_{\min}$, and
\item $c_2$ is represented by the path $\gamma'_{\min}, \gamma_0, \ldots, \gamma_n, \gamma'_{\max}$.
\end{itemize}

Clearly, in this construction, the graph $H_3$ is correctly represented.
Moreover, the subtree corresponding to every minimal element includes all of the nodes $v_{\min},$
$\alpha_{\min},$ $\alpha'_{\min},$ $\beta_{\min},$ $\beta'_{\min},$ $\gamma_{\min},$
$\gamma'_{\min}$, but none of the opposite $\max$-nodes. 
Thus, each minimal element is
universal to $V_{\min}$ and non-adjacent to the vertices of $V_{\max} \setminus \{a_2, b_2, c_3\}$, as needed. 
Symmetrically,
each maximal element is universal to $V_{\max}$ and non-adjacent to the vertices of $V_{\min} \setminus \{a_2, b_2, c_3\}$.
It follows that $G$ is an $H$-graph. 
\end{proofclaim}

This completes the first part of the proof.

\heading{Part 2: $H$ contains the diamond graph $D$ as a minor.}
The argument here follows very similarly to the proof shown in Part~1.
We again use the 3-subdivision $H_3$ of $H$ which, by Lemma~\ref{lem:3-subdivision-blocker}, canonically ``covers'' $H$. 
Again, $H_3$ will be used as part of the graph $G$ we will construct from $G_\calP$ so that $G \in \graphs{H}$ if and only if $\calP$ has interval dimension at most 3. 
Importantly, $H_3$ also allows us to, with a careful choice of $V_{\min}$ and $V_{\max}$, appropriately restrict the representations of the minima and maxima to only use a chosen diamond minor of $H$ (up to automorphism of course) as before.

Observe that, since $H$ contains $D$ as a minor (and the maximum degree of $D$ is three), a subdivision of $D$ (or $D$ itself) is a subgraph of $H$. Let $D^*$ be a subgraph of $H$ that is a subdivision of $D$. 
In particular, $D^*$ consists of two nodes $d_{\min}$ and $d_{\max}$ of degree 3 and three $(d_{\min},d_{\max})$-paths $A,B,C$ that are edge disjoint and whose internal vertices are of degree~$2$.
Let $\alpha = d_{\min}d^{\alpha}_{\max} \in A$, $\beta = d_{\min}d^{\beta}_{\max} \in B$, and $\gamma = d_{\min}d^{\gamma}_{\max} \in C$ denote the three edges incident to the node $d_{\min}$ in $D^*$. 
These three edges will be used equivalently to the three $v_{\min}v_{\max}$ edges as in Part~1, i.e., they will the ``location'' in $H$ where we will see the three intervals certifying that our original poset has interval dimension~$3$.

Now, let $D^*_3$ be the subgraph of $H_3$ corresponding to $D^*$ ($D^*_3$ is also a subdivision of $D$). 
Let $z_{\min}, z_{\max}$ 
be the vertices in $D^*_3$ corresponding (via the subdivision of $H$ to $H_3$) to $d_{\min}, d_{\max}$ 
 in $D^*$ respectively, and further:
\begin{itemize}
\item let $z_{\min}, a_1, a_2, a_3, a_{\max}$ be the path in $D^*_3$  corresponding to the subdivision of the edge $\alpha$ of $D^*$, and 
\item let $z_{\min}, b_1, b_2, b_3, b_{\max}$ be the path in $D^*_3$  corresponding to the subdivision of the edge $\beta$ of $D^*$, and 
\item let $z_{\min}, c_1, c_2, c_3, c_{\max}$ be the path in $D^*_3$  corresponding to the subdivision of the edge $\gamma$ of $D^*$. 
\end{itemize}

We are now ready to construct our graph $G$ from $H_3$ and the graph $G_\calP$ of a given height one poset $\calP=(P,<)$ so that $G \in \graphs{H}$ if and only if $\calP$ has interval dimension three.
Recall that $K_{\min}$ and $K_{\max}$ denote cliques on the minima and maxima  of $\calP$ respectively. 
As in Part~1, we let $V_{\min} = \{z_{\min}, a_1, a_2, b_1, b_2, c_1, c_2\}$. 
Similarly to Part~1, we let $V_{\max}$ be the vertex set of the minimal subgraph of $D^*_3$ containing $\{z_{\max}, a_2, b_2, c_2\}$. 
In other words $V_{\max} = V(D^*_3) \setminus V_{\min} \cup \{a_2, b_2, c_2\} = V(D^*_3) \setminus \{z_{\min}, a_1, b_1, c_1\}$.
Now, as in Part~1, the graph $G$ is the union of $G_\calP$ and $H_3$ where, additionally, each vertex of $K_{\min}$ is adjacent to each vertex of $V_{\min}$ and each vertex of $K_{\max}$ is adjacent to each vertex of $V_{\max}$. 

The completion of the proof now follows nearly identically to the proof of the claim in Part~1. 
Namely, by Lemma~\ref{lem:3-subdivision-blocker}, $H_3$ has a unique up to automorphism $H$-representation, and the vertices of $K_{\min}$ and $K_{\max}$ can essentially only be represented on the $D^*$ part of $H$ (due to their adjacency with the vertices of $H_3$)\footnote{While the representation of a vertex of $G_\calP$ might ``reach out'' beyond $D^*$ onto an incident edge, it can never traverse all of such an edge because, by Lemma~\ref{lem:3-subdivision-blocker}, there is a vertex $x_e$ of $H_3$ occupying the ``middle'' of that edge and, by construction, $x_e$ is not adjacent to any vertex of $G_\calP$.}. 
Moreover, within the three edges $\alpha, \beta, \gamma$, there will be the representations of $a_2, b_2, c_2$ and within these representations we will indeed have the (at most) $3$ interval models. 
\end{proof}

The next section gives a positive answer for the following problem in the case when $H$ is a tree.
Also, recall that when $H$ is a single cycle, $\graphs{H}$ is the class of circular-arc graphs and as such can be recognized in linear time. 
This leaves the following problem. 

\begin{problem}
For a non-tree fixed cactus graph $H$ (other than a single cycle), is there a polynomial-time time algorithm testing whether $G$ is an $H$-graph?
\end{problem}

\section{Polynomial-time recognition algorithms}
\label{sec:recog-poly}

We present an $\calO(n^{3.5})$-time algorithm recognizing $S_d$-graphs and an $\cXP$-time algorithm
recognizing $T$-graphs (parametrized by the size of the tree $T$).  We begin with a lemma that
motivates our approach. It implies that if $G$ is a $T$-graph, then there exists a
representation of $G$ such that every branching point is ``contained" in some maximal clique of $G$.

\begin{lemma}\label{lem:clique_branching_point}
For any $T$-graph $G$ and $T$-representation $\calR$ of $G$, 
$\calR$ can be modified such that for every branch node $b \in V(T')$, we have $b \in \bigcap_{v \in C}V(T_v')$, for some maximal clique
$C$ of $G$.
\end{lemma}

\begin{proof}
For every node $x$ of the subdivision $T'$, let $V_x = \{u \in V(G) : x \in V(T_u')\}$ be the set of
vertices of $G$ corresponding to the subtrees passing through $x$. Let $b$ be a branching point such
that $V_b$ is not a maximal clique.

We pick a maximal clique $C$ with $C \supsetneq V_b$. Since
$\calR$ satisfies the Helly property, there is a node $a \in \bigcap\{V(T_v') : v \in C\}$.
Note that for every node $x$ of $P_{[a,b]}$, we have $V_x \supseteq V_b$.
Let $x$ be the node of $P_{(b,a]}$ closest to $b$ such that $V_x\neq V_b$.
Then, for each $v \in V_x
\setminus V_b$, we update $T_v'$ to be $T_v' \cup P_{[b,x]}$. Thus, we obtain a correct
representation of $G$ with $V_b = V_x$.

We repeat the process described in the previous paragraph until $V_b$ is a maximal clique.
\end{proof}

\paragraph{Remark on subdivisions.}
For convenience, we assume throughout the whole section that we already have a sufficiently large
subdivision $T'$ of $T$. At the end, it will be clear that a subdivision $T'$ of $T$ with $|V(T')|
\leq cn + |V(T)|$, for some constant $c$, suffices. In fact, it suffices to have $c=3$.

\paragraph{General idea.}
It is well-known that chordal graphs, and therefore also $T$-graphs, have at most $n$ maximal
cliques and that they can be listed in linear time. Let $\calB$ be the set of branching points of $T$
and let $\calC$ be the set of all maximal cliques of $G$.
The main part of our algorithm attempts, for a given $f \colon\calB\to\calC$, to construct a
$T$-representation satisfying $V_b = \bigcap_{v \in f(b)}V(T_v')$, for every $b \in \calB$, where $V_b = \{u \in V(G) : b \in V(T_u')\}$.
By Lemma~\ref{lem:clique_branching_point}, there always exists such a representation.


To this
end, we try find interval representations of the connected components of $G - \bigcup_{b \in
\calB}f(b)$ on the paths $T' - \calB$ such that the following
conditions hold:
\begin{itemize}
\item[(i)]
If interval representations of the connected components $X_1,\dots, X_k$ are on a path $P_{(b,l]}$,
where $b \in\calB$ and $l$ is a leaf of $T'$, then the induced subgraph $G[f(b)\cup
V(X_1)\cup\cdots\cup V(X_k)]$ has an interval representation on $P_{[b,l]}$ in which $f(b)$ is the leftmost clique.

\item[(ii)]
If interval representations of the connected components $X_1,\dots, X_k$ are on a path $P_{(b,b')}$,
where $b,b' \in \calB$, then the induced subgraph $G[f(b)\cup V(X_1)\cup\cdots\cup
V(X_k)\cup f(b')]$ has an interval representation on $P_{[b,b']}$ in which $f(b)$ and $f(b')$ are the rightmost and leftmost cliques, respectively.
\end{itemize}
%
%

\subsection{Recognition of \texorpdfstring{$S_d$}{Sd}-graphs}

In the case when $T = S_d$, we have $\calB = \{b\}$ and $V(T) = \{b\} \cup \{l_1,\dots,l_d\}$.  The
number of mappings $f:\{b\}\to\calC$ is exactly the same as the number of maximal cliques of $G$,
which is at most $n$ (otherwise it is not an $S_d$-graph).  For every maximal clique $C$ of $G$, we
try to construct a $T$-representation $\calR$ such that $b \in \bigcap_{v \in C}V(T_v')$.

\begin{figure}[t]
\centering
\includegraphics{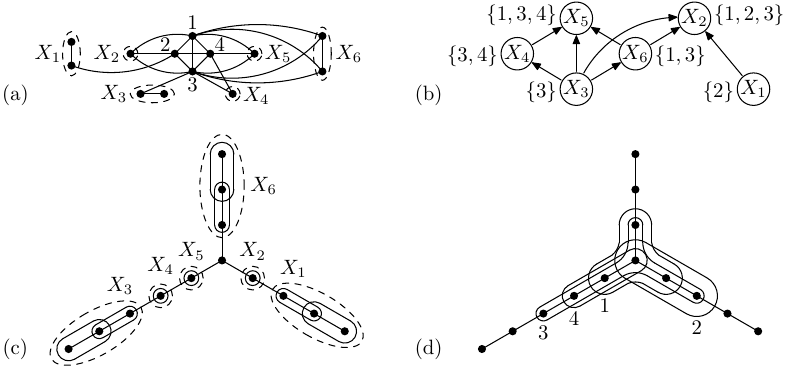}
\caption{
(a) An example of an $S_d$-graph $G$ with a maximal clique $C = \{1,2,3,4\}$. (b) The partial
ordering $\triangleright$ on the connected components of $G - C$ with chain cover of size $3$:
$X_2\triangleright X_1$, $X_5\triangleright X_4\triangleright X_3$, and $X_6$. (c) The connected
components placed on the paths $P_{(b,l_1]}$, $P_{(b,l_2]}$, and $P_{(b,l_3]}$, according to the
chain cover of $\triangleright$. (d) The subtrees $T_1', T_2', T_3', T_4'$ corresponding to the
vertices of the maximal clique $C$ give an $S_d$-representation of $G$ with $b \in \bigcap_{v\in
C}V(T_v')$.}
\label{fig:star_rep_construction}
\end{figure}

Assume that $G$ has such an $S_d$-representation, for some maximal clique $C$. Then the connected
components of $G - C$ are interval graphs and each connected component can be represented on one of
the paths $P_{(b,l_i]}$, which is a subdivision of the edge $bl_i$; see
Fig.~\ref{fig:star_rep_construction}a and~\ref{fig:star_rep_construction}c. However, some pairs of
connected components of $G - C$ cannot be placed on the same path $P_{(b,l_i]}$, since their
``neighborhoods'' in $C$ are not ``compatible''.  The idea is to define a partial order
$\triangleright$ on the components of $G - C$ such that for every linear chain
$X_1\triangleright\dots\triangleright X_k$, the induced subgraph  $G[C, V(X_1),\dots, V(X_k)]$ can
be represented on some path $P_{(b,l_i]}$; see Fig.~\ref{fig:star_rep_construction}b.

We define $N_C(u)$ and $N_C(X)$ to be the \emph{neighbourhoods} of the vertex $u$ in $C$ and of the
components $X$ in $C$, respectively. Formally,
$$N_C(u) = \{v \in C : vu\in E(G)\}\quad\text{and}\quad N_C(X) = \bigcup\{N_C(u) : u\in V(X)\}.$$
Note that, if we have two components $X$ and $X'$ on the same branch where $N_C(X') \subseteq
N_C(u)$ for every $u \in V(X)$, then $X$ must be closer to $C$ than $X'$ if they are represented on the same path $P_{(b,l_i]}$. 

We say that components $X$ and $X'$ are \emph{equivalent}, $X\sim X'$, if there
is a subset $C'$ of $C$ such that $N_C(u) = C'$ for every $u \in V(X)$ and $N_C(u') = C'$ for every
$u' \in V(X')$.  Note that equivalent components $X$ and $X'$ can be represented in an interval
representation of $G[C,V(X),V(X')]$ in an arbitrary order and they \emph{can be treated as one
component}. We denote the set of the equivalence classes $G-C/\sim$ by $\calX$. For $X, X' \in \calX$, we put:
\begin{equation}\label{eq:poset_condition}
X \triangleright X' \quad \text{if for every } u \in V(X), N_C(X') \subseteq N_C(u) \text{ or if } X = X'.
\end{equation}

\begin{lemma}\label{lem:partial_ordering}
The relation $\triangleright$ is a partial ordering on $\calX$.
\end{lemma}

\begin{proof}
The relation $\triangleright$ is reflexive by definition.  Suppose that $X \triangleright X'$ and $X'
\triangleright X$. For every $u \in V(X)$ and $u' \in V(X')$, we have
$$N_C(u') \subseteq N_C(X') \subseteq N_C(u)\quad\text{and}\quad N_C(u)\subseteq N_C(X)\subseteq
N_C(u').$$
Therefore, $N_C(u) = N_C(u')$ for every $u\in V(X)$ and $u' \in V(X')$ and $X$ and $X'$ are
equivalent. We assume that $\calX$ contains only non-equivalent components. So, $X = X'$ and the
relation $\triangleright$ is asymmetric. It can be easily checked that $\triangleright$ is also
transitive.
\end{proof}

\begin{lemma}\label{lem:branch_representation}
Let $X_1, \dots, X_k \in \calX$. Then the induced subgraph $G[C, V(X_1), \dots, V(X_k)]$ has an
interval representation with $C$ being the leftmost clique if and only if $X_1\triangleright
\dots\triangleright X_k$ and each $G[C,X_i]$ has an interval representation with $C$ being the
leftmost clique.
\end{lemma}

\begin{proof}
Suppose that there is an interval representation $\calR$ of $G[C, V(X_1),\dots, V(X_k)]$ with $C$
being the leftmost maximal clique. Since each $X_i$ is a connected components of $G - C$, their
representations in $\calR$ cannot overlap. Without loss of generality, we assume that the components
$X_1, \dots, X_k$ are ordered such that $i<j$ if and only if $X_i$ is placed closer to $C$ in
$\calR$ than $X_j$. Let $u \in V(X_i)$ and $v \in N_C(X_j)$. The vertex $v$ is adjacent to at least
one vertex of $X_j$. Therefore, the representation of $v$ covers the whole component $X_i$ in
$\calR$, i.e., we have $v \in N_C(u)$ and $X_i \triangleright X_j$.

For the converse, we assume that $X_1, \dots, X_k$ form a chain in $\triangleright$ and every $G[C,
X_i]$ has an interval representation $\calR_i$ with $C$ being the leftmost clique. Since $X_i
\triangleright X_j$, for $i<j$, every vertex in $N_C(X_j)$ is adjacent to every vertex of $X_i$.
We now construct an interval representation of $G[C, V(X_1), \dots, V(X_k)]$. We first place the interval
representations of all $X_i$'s (i.e., we use $\calR_i$ restricted to the intervals of $V(X_i)$) on the 
real line according to $\triangleright$, with $X_1$ being the leftmost. 
Let $x_1,\dots,x_{k+1} \in \mathbb{R}$ be the points of the real line such that $X_i$ is
represented on the interval $(x_i,x_{i+1})\subseteq\mathbb{R}$. 

It remains to construct a representation for every vertex $v \in C$. Let
$$C_k = N_C(X_k) ~\text{and}~ C_i = N_C(X_i) \setminus \bigcup_{j=i+1}^{k} N_C(X_{j}), i =
0,\dots,k-1 ~\text{where}~ X_0 = C.$$
Let $x_0 \in \mathbb{R}$ be a point left of $x_1$. All the vertices in $C_0$ are represented by the
interval $[x_0,y]$, for some $y < x_1$. The intervals representing vertices in $C_i$ are constructed
inductively, for $i = k, k-1,\dots,1$. For $i \leq k$, we assume that we constructed the
representations of vertices in $C_{i+1},\dots,C_k$. Note, if  $X_j \triangleright X_i$, then for
every $u \in V(X_j)$, we have $N_C(X_i) \subseteq N_C(u)$. Therefore, every vertex in $C_i$ is
represented by an interval of the form $[x_0,z]$, where $z \in (x_i,x_{i+1})$ is a suitable point
given by the representation $\calR_i$ of $G[C,X_i]$.
\end{proof}

The following theorem gives a characterization of $S_d$-graphs.
It generalizes the characterization of interval graphs due to Fulkerson and Gross; see Lemma~\ref{lem:fulkerson_gross}.

\begin{theorem}[Characterization of $S_d$-graphs]\label{thm:star_representation}
A graph $G$ is an $S_d$-graph if an only if there is a maximal clique $C$ of $G$ such that the following hold:
\begin{itemize}
\item[(i)]
For every connected component $X$ of $G-C$, the induced subgraph $G[C,X]$ has an interval representation with $C$ being the leftmost clique.
\item[(ii)]
The partial order $\triangleright$ on $\calX = G - C/\sim$ has a chain cover of size at most $d$.
\end{itemize}
\end{theorem}

\begin{proof}
Suppose that $G$ is an $S_d$-graph with a representation satisfying $b \in \bigcap_{v \in
C}V(T_v')$; such a representation always exists by Lemma~\ref{lem:clique_branching_point}. 
The representation of a connected component $X \in \calX$ can not pass through the node
$b$ since otherwise $C$ would not be a maximal clique. Clearly, the conditions (i) is satisfied. The
representations of every two components in $\calX$ have to be placed on non-overlapping parts of the
subdivided $S_d$. By Lemma~\ref{lem:branch_representation}, we have that the components placed on
some path $P_{(b,l_i]}$ of the subdivided $S_d$ form a linear chain in $\triangleright$. Therefore,
the partial order $\triangleright$ has a chain cover of size at most $d$ and the condition (ii) is
satisfied; see Fig.~\ref{fig:star_rep_construction}b.

Suppose that the conditions (i) and (ii) are satisfied. We put the components in $\calX$ on the paths
$P_{(b,l_1]}, \dots, P_{(b,l_d]}$ according to the chain cover of the partial order $\triangleright$
which has size at most $d$, i.e, every chain of $\triangleright$ is placed on one $P_{(b,l_i]}$. By
Lemma~\ref{lem:branch_representation}, for every chain $X_1, \dots, X_k$ in $\triangleright$, we can
find an interval representation of the graph $G[C, V(X_1), \dots, V(X_k)]$ with $C$ being the
leftmost maximal clique.
\end{proof}

\paragraph{Algorithm.}
By combining Lemmas~\ref{lem:branch_representation} and Theorem~\ref{thm:star_representation} we obtain an
algorithm for recognizing $S_d$-graphs. For a given graph $G$ and its maximal clique $C$, we do the
following:
\begin{enumerate}
\item
We delete the maximal clique $C$ and construct the partial order $\triangleright$ on the set of
non-equivalent connected components $\calX$.
\item
We test whether the partial order $\triangleright$ can be covered by at most $d$ chains.
\item
For each linear chain $X_1^i\triangleright\cdots\triangleright X_k^i$, $1\leq i\leq d$, we construct
an interval representation $\calR_i$ of the induced subgraph $G[C,V(X_1^i),\dots,V(X_k^i)]$, with $C$
being the leftmost maximal clique, on one of the paths of the subdivided $S_d$.
\item 
We complete the whole representation by placing each $\calR_i$ on the path $P_{[b,l_i]}$ so that $b
\in \bigcap_{v \in C}V(T_v')$. 
\end{enumerate}

\begin{theorem}\label{thm:poly_star}
Recognition of $S_d$-graphs can be solved in $\calO(n^{3.5})$ time.
\end{theorem}

\begin{proof}
Every chordal graph has at most $n$ maximal cliques, where $n$ is the number of vertices, and they
can be listed in linear time~\cite{rose1976algorithmic}. For every clique $C$, our algorithm tries
to find an $S_d$-representation with $b \in \bigcap_{v\in C}V(T_v')$. The partial order $\triangleright$ can be constructed in time $\calO(n^2)$. By forgetting the
orientation in the partial order $\triangleright$, we get a comparability graph, and every clique in
the comparability graph induces a linear chain in $\triangleright$. A relatively simple algorithm
finds a minimum
clique-cover of a comparability graph in time $\calO(n^3)$~\cite{golumbic1977complexity}. An
algorithm that runs in time $\calO(n^{2.5})$ can by obtained by a combination
of~\cite{fulkerson1956note} and~\cite{hopcroft1973n}.
Testing whether $G[C,V(X_1^i),\dots,V(X_k^i)]$ has an interval representation with $C$ being the leftmost maximal clique can be done in linear time.
Thus, the overall time
complexity of our algorithm is $\calO(n^{3.5})$.
\end{proof}

\begin{problem}
Can we recognize $S_d$-graphs in time $\calO(n^{2.5})$? In particular, can we find the clique that can
be placed in the center of $S_d$ efficiently?
\end{problem}

\subsection{Recognition of T-graphs}

The algorithm for recognizing $T$-graphs is a generalization of the algorithm for recognizing
$S_d$-graphs described above. Let $f\colon \calB\to\calC$ be an fixed assignment of cliques. 

\paragraph{Assumption (connectedness of $\boldsymbol{G}$).}
Suppose that $G$ is disconnected. Then it can be written as a disjoint union of some $X$ and
$\widehat{G}$, where $X$ is a connected component of $G$. Let $\calC_X$ and $\widehat{\calC}$ be the
maximal cliques of $X$ and $\widehat{G}$, respectively. The sets $f^{-1}(\calC_X)$ and
$f^{-1}(\widehat{\calC})$ induce subtrees $T_X$ and $\widehat{T}$ of $T$ separated by the branch $ab$,
where $a \in V(T_X)$ and $b \in V(\widehat{T})$ (otherwise $f$ is invalid). We subdivide the branch
$ab$ by nodes $c_1$ and $c_2$. Then we try to find a representation of $X$ on the tree $T_X \cup
ac_1$ and a representation of $\widehat{G}$ on $\widehat{T} \cup c_2b$. Therefore, we may assume
that $G$ is connected.

\paragraph{Assumption (injectiveness of $\boldsymbol{f}$).}
Suppose that $f$ is not injective, i.e., $f(b) = f(b')$.  Then for every branching point $b''$ which
lies on the path from $b$ to $b'$, we must have $f(b) = f(b'') = f(b')$ (otherwise $f$ is invalid).
For $C \in f(\calB)$, the branching points in $f^{-1}(C)$, together with the paths connecting them,
have to form a subtree $T_C$ of $T$. In this case the whole subtree $T_C$ can be contracted into a
single node $a$.
Note that if there is a component $X$ of $G-\bigcup_{b\in \calB}f(b)$ where every vertex of $X$ adjacent to every vertex of $C$,
then $X$ can be represented on any branch incident to $a$ by subdividing it appropriately.
Thus, we may assume that $f$ is injective.

\paragraph{Step 1 (components between branching points).}
The first step of our algorithm is to find for $b,b'\in E(T)$, which components \emph{have to} be
represented on the path $P_{(b, b')}$ of $T'$.

\begin{lemma}\label{lem:components_middle}
Let $X$ be a connected component of $G - \bigcup_{b \in \calB}f(b)$ and $bb' \in E(T)$. If the
sets
$$(f(b)\setminus f(b')) \cap N_{f(b)}(X) \neq \emptyset \quad\text{and}\quad (f(b')\setminus f(b))
\cap N_{f(b')}(X) \neq \emptyset,$$
then $X$ has to be represented on $P_{(b,b')}$ of $T'$.
\end{lemma}

\begin{figure}[t]
\centering
\includegraphics{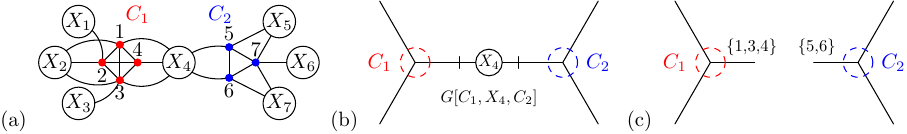}
\caption{(a) A $T$-graph $G$, where $T$ is the tree shown on the right. We have $f(b) = C_1$ and
$f(b') = C_2$. (b) Component $X_4$ with $C_1\setminus C_2 \cap N_{C_1}(X) \neq
\emptyset$ and $C_2\setminus C_1 \cap N_{C_2}(X) \neq \emptyset$. In this case, $X_{b,b'} = X_4$.
(c) A segment of the star corresponding to the clique is labeled by $\{1,3,4\} \subseteq C_1$. A
component $X$ can be represented on this segment only if $\{1,3,4\} \subseteq N_{C_1}(X)$.}
\label{fig:generalt}
\end{figure}

\begin{proof}
Let $v \in (f(b)\setminus f(b')) \cap N_{f(b)}(X)$ and $u \in (f(b')\setminus f(b)) \cap
N_{f(b')}(X)$.  Since $v \notin f(b')$, we have $b' \notin V(T_v')$. Similarly we have $b \notin
V(T_u')$.
Putting it together, we have that $b \in V(T_v')$ and $b'\notin V(T_v')$, and $b \notin V(T_u')$ and $b'\in V(T_u')$.
Since $X$ is adjacent to both $u$ and $v$, the only possible path where $X$ can be represented is
$P_{(b,b')}$; see Fig.~\ref{fig:generalt}a and~\ref{fig:generalt}b.
\end{proof}

We do the following for each $b,b' \in \calB$ such that $bb' \in E(T)$.  Let $X_{b,b'}$ be the
disjoint union of the components satisfying the conditions of Lemma~\ref{lem:components_middle}. If
the induced interval subgraph $G[C \cup V(X_{b,b'}) \cup C']$ has a representation such that the
cliques $C$ and $C'$ are the leftmost and the rightmost, respectively, then we can represent
$X_{b,b'}$ in the middle of the path $P_{(b,b')}$. If no such representation exists, then $G$ does
not have $T$-representation for this particular $f \colon \calB \to \calC$.  This means that
the representation of $X_{b,b'}$ is constructed on a proper subpath of $P_{(b,b')}$ -- recall that we are assuming that the subdivision $T'$ is sufficiently large.

Next, we do the following for every $b \in \calB$. Let $l_1,\dots,l_p$ and $b_1,\dots,b_q$ be the
leaves of $T$ and the branching points of $T$, respectively, such that $bl_i \in E(T)$, for every $i
= 1,\dots,p$, and $bb_j \in E(T)$, for every $j = 1,\dots,q$. Let $a_1,\dots,a_q$ and
$a'_1,\dots,a'_q$ be the points of the paths $P_{[b,b_1]}, \dots, P_{[b,b_q]}$, respectively, such
that $X_{b,b_i}$ is represented on the subpath $P_{(a_i,a'_i)}$. We define $S(b)$ to be the subdivided
star consisting of the paths $P_{[b,l_1]}, \dots,P_{[b,l_p]}, P_{[b,a_1)},\dots,P_{[b,a_q)}$.
Note that if a vertex $u \in V(X_{b,b_i})$ is adjacent to a vertex $v$ in $f(b)$, then the
representation of $T_v'$ of $v$ contains the whole subpath $P_{(b,a_i)}$. This means that a
component $X$, which does not satisfy the condition of Lemma~\ref{lem:components_middle},  can be represented on $P_{(b,a_i)}$ only if $N_{f(b)}(X_{b,b_i}) \subseteq
N_{f(b)}(X)$. We remove the subpaths $P_{(a_i,a_i')}$ (together with the representations of $X_{b,b_i'}$)
and we are left with disjoint subdivided stars with restrictions; see Fig~\ref{fig:generalt}c.

\paragraph{Step 2 (disjoint stars with restrictions).}
We reduced the problem of recognizing $T$-graphs to the following problem.  Let $H$ be a fixed graph
formed by the disjoint union of $k$ stars $S(b_1),\dots,S(b_k)$ with branching points
$b_1,\dots,b_k$.  On the input we have a graph $G$, an injective mapping $f \colon \{b_1,\dots,b_k\}
\to \calC$, and for every edge of $S(b_i)$ a subset of $f(b_i)$, called \emph{restrictions}. We want
to find a representation of $G$ on $H$ such that $b_i \in \bigcap_{v\in f(b_i)}V(H_v')$, and for
every connected component $X$ of $G - \bigcup_{i=1}^k f(b_i)$, the vertices $V(X)$ have to
be adjacent to every vertex in the restrictions corresponding to the path on which $X$ is
represented.

To solve this problem, we define a partial ordering on the connected components of $G - C$, where $C
= \bigcup f(b_i)$. The notions $N_C(u)$ and $N_C(X)$ are defined as in the same way as in the
algorithm for recognizing $S_d$-graphs.  We get a partial ordering $\triangleright$ on the set of
non-equivalent connected components $\calX$ of $G - C$. Moreover, to each component $X \in \calX$,
we assign a list of colors $L(X)$ which correspond to the subpaths from a branching point to a leaf
in the stars $S(b_1), \dots, S(b_k)$, on which they can be represented. Each list $L(X)$ has size at
most $d = \sum_{i=1}^k d_i$, where $d_i$ is the degree of $b_i$.

Suppose that there exists a chain cover of $\triangleright$ of size $d$ such that for every chain
$X_1\triangleright\dots\triangleright X_\ell$ in this cover we can pick a color belonging to every
$\bigcap_{j = 1}^\ell L(X_j)$ such that no two chains get the same color. In that case a
representation of $G$ satisfying the restrictions can be constructed analogously as in the proof of
Lemma~\ref{lem:branch_representation} and~\ref{thm:star_representation}.

The partial ordering $\triangleright$ on the components $\calX$ defines a comparability graph $P$
with a list of colors $L(v)$ assigned to every vertex $v\in V(P)$. If we find a \emph{list
coloring} $c$ of its complement $\overline{P}$, i.e., a coloring that for every vertex $v$ uses only
colors from its list $L(v)$, then the vertices of the same color in $\overline{P}$ correspond to a
chain (clique) in $P$. Therefore, we have reduced our problem to list coloring co-comparability
graphs with lists of bounded size.

\paragraph{Step 3 (bounded list coloring of co-comparability graphs).}
We showed that to solve the problem of recognizing $T$-graphs we need to solve the $\ell$-list
coloring problem for co-comparability graphs where $\ell = 2\cdot|E(T)|$. In particular, given a
co-comparability graph $G$, a set of colors $S$ such that $|S| \leq \ell$, and a set $L(v) \subseteq
S$ for each vertex $v$, we want to find a proper coloring $c \colon V(G) \to S$ such that for every
vertex $v$, we have $c(v) \in L(v)$.

In~\cite{bonomo2011bounded}, the authors consider the \emph{capacitated coloring problem} for
co-comparability graphs. Namely, given a graph $G$, an integer $s \geq 1$ of colors, and positive
integers $\alpha_1^*, \dots,\alpha_s^*$, a \emph{capacitated $s$-coloring $c$ of $G$} is a proper
$s$-coloring such that the number of vertices assigned color $i$ is bounded by $\alpha_i^*$, i.e.,
$|c^{-1}(i)| \leq \alpha_i^{*}$. The authors prove that the capacitated coloring of co-comparability
graphs can be solved in polynomial time for fixed $s$. In the next section, we modify their approach
to solve the $s$-list coloring problem on co-comparability graphs in $\calO(n^{s^2 + 1}s^3)$ time.
This provides the following theorem.

\begin{theorem}\label{thm:xp_tree}
Recognition of $T$-graphs can be solved in $n^{\calO(\|T\|^2)}$.
\end{theorem}

\begin{problem}
Is there an \cFPT\ algorithm for recognizing $T$-graphs?
\end{problem}

\subsection{Bounded list coloring of co-comparability graphs}\label{sec:list_col}

Here, we provide a polynomial time algorithm for the problem of bounded list coloring of co-comparability graphs. 
Our result can be seen as a generalization
of the polynomial time algorithm of Enright, Stewart, and 
Tardos~\cite{EnrightST14} for bounded list coloring on a class which includes both interval graphs and permutation graphs. 
However, they~\cite{EnrightST14} explicitly state that their approach does not extend to co-comparability graphs. 
To prove this also for co-comparability graphs, we slightly modify the approach in~\cite{bonomo2011bounded}. 

In~\cite{bonomo2011bounded}, the problem of capacitated coloring is solved for a
more general class of graphs, so called $k$-thin graphs. A graph $G$ is
\emph{$k$-thin} if there exists an ordering $v_1,\dots,v_n$ of $V(G)$ and a
partition of $V(G)$ into $k$ classes $V^1,\dots,V^k$ such that, for each triple
$p,q,r$ with $p<q<r$, if $v_p,v_q$ belong to the same class and $v_rv_p \in
E(G)$, then $v_rv_q \in E(G)$. Such ordering and partition are called
\emph{consistent}. The minimum $k$ such that $G$ is $k$-thin is called the
\emph{thinness} of $G$. Graphs with bounded thinness were introduced
in~\cite{mannino2007stable} as a generalization of interval graphs. Note that
interval graphs are exactly the $1$-thin graphs.

Recall that a graph $G$ is a \emph{comparability graph} if there exits an
ordering $v_1,\dots,v_n$ of $V(G)$ such that, for each triple $p,q,r$ with
$p<q<r$, if $v_pv_q$ and $v_qv_r$ are edges of $G$, then so is $v_pv_r$. Such an
ordering is a \emph{comparability ordering}.

\begin{lemma}[Theorem 8, \cite{bonomo2011bounded}]
\label{thm:coco_kthin}
Let $G$ be a co-comparability graph. Then the thinness of $G$ is at most
$\chi(G)$, where $\chi$ is the chromatic number. Moreover, any vertex partition
given by a coloring of $G$ and any comparability ordering for its complement are
consistent.
\end{lemma}

Let $G$ be $k$-thin graph, and let $v_1,\dots,v_n$ and $V^1,\dots,V^k$ be an
ordering and a partition of $V(G)$ which are consistent. Note that the ordering
induces an order on each class $V^j$. For each vertex $v_r$ and class $V^j$, let
$N(v_r,j)_<$ be the set of neighbors of $v_r$ in $V^j$ that are smaller than
$v_r$, i.e., $N(v_r,j)_< = V^j \cap \{v_1,\dots,v_{r-1}\}\cap N(v_r)$. For each
class $V^j$ let $\Delta(j)_<$ be the maximum size of $N(v_r,j)_<$ over all
vertices $v_r$. The following lemma gives an alternative definition of $k$-thin
graphs.

\begin{lemma}[Fact 7, \cite{bonomo2011bounded}]
For each vertex $v_r \in \{v_1,\dots,v_n\}$ and each $j \in
\{1,\dots,k\}$, the set $N(v_r,j)_<$ is such that:
\begin{itemize}
\item
the vertices in $N(v_r,j)_<$ are consecutive, with respect to the order induced
on $V^j$.
\item
if $N(v_r,j)_< \neq \emptyset$, then it includes the vertex with largest index in $V^j
\cap\{v_1,\dots,v_{r-1}\}$.
\end{itemize}
\end{lemma}

\paragraph{Bounded List Coloring On $k$-thin Graphs.}
In~\cite{bonomo2011bounded}, the problem of capacitated coloring is reduced to a
reachability problem on an auxiliary acyclic digraph.\footnote{Note that this is just a representational convenience for dynamic programming.} We obtain an algorithm
for bounded list coloring on $k$-thin graphs by slightly modifying the algorithm
for capacitated coloring in~\cite{bonomo2011bounded}. The only difference is
that we do not have a restriction on how many times we can use a particular
color and for every vertex we can only use the colors from the list assigned to
it.  Otherwise, everything is the same as in~\cite{bonomo2011bounded}. We
include it here for the sake of completeness.

Let $G$ be a $k$-thin graph with an ordering $v_1,\dots,v_n$ and a partition
$V^1,V^2,\dots,V^k$ of $V(G)$. Let $S$ be a set of colors, $s = |S|$, and $L:V(G)\to
\mathcal{P}(S)$ be a function that assigns a list of allowed colors to a vertex.
Consider an instance $(G,L)$ of list coloring. We reduce the problem to a
reachability problem on an auxiliary acyclic digraph $D(N,A)$. We will refer to the
elements of $N$ and $A$ as \emph{nodes} and \emph{arcs} while the elements of
$V(G)$ and $E(G)$ will be referred to as \emph{vertices} and \emph{edges} (as we did
so far).

The digraph $D$ will be \emph{layered}, i.e., the set $N$ is the disjoint union
of subsets (layers) $N_0, N_1,\dots,N_n$ and all arcs in $A$ have the form
$(u,w)$ with $u\in N_r$ and $w\in N_{r+1}$, for some $0\le r \le n-1$. Note for each vertex $v_r \in V$, there is a layer $N_r$ with $r\neq 0$. 
We denote by $j(r)$ the class index $q$ such that $v_r\in V^q$.

We first describe the set of nodes in each layer. The first layer consists of
colors which can be assigned to the first vertex, i.e.,  $N_0=L(v_1)$. For the
layers $N_1,\dots,N_{n-1}$, there is a one-to-one correspondence between nodes at
layer $N_r$ and $(sk+1)$-tuples $(r,\{\beta_i^j\}_{i=1,\dots,s,j=1,\dots,k})$ with $0\le
\beta_i^j\le \Delta(j)_<$, for each $i,j$. The last layer $N_n$ has only
one node $t$ corresponding to the tuple $(n,0,\dots,0)$.

We associate with each node $u\notin N_0$ a suitable list coloring problem
\emph{with additional constraints}, that we call the \emph{constrained sub-problem}
associated with $u$. As we show in the following, $u$ is reachable from a node
$z\in N_0$ if and only if this constrained sub-problem has a solution. Namely,
we will show that the following property holds:
\begin{enumerate}
\renewcommand{\theenumi}{$(*)$}
\renewcommand{\labelenumi}{$(*)$}
\item\label{prop} a node $(r,\{\beta_i^j\}_{i=1,\dots,s,j=1,\dots,k})$ is reachable from a
node $z\in N_0$ if and only if the induced subgraph $G[\{v_1,\dots,v_r\}]$ admits a list
coloring with the lists given by $L$ and with additional constraint
that, for each $i=1,\dots,s$ and $j=1,\dots,k$, color $i$ is forbidden for the
last $\beta_i^j$ vertices in $V^j\cap \{v_1,\dots,v_r\}$.
\end{enumerate}
In this case, $G$ admits a list coloring if and only if the node $t$ is
reachable from a node $z\in N_0$.

Property~\ref{prop} will follow from the definition of the set of arcs $A$ given
as follows. Let $u=(r,\{\beta_i^j\}_{i=1,\dots,s,j=1,\dots,k})$.
Note that the problem associated with $u$ has a solution where the vertex $v_r$
gets color $i$ only if $\beta_i^{j(r)}=0$. Let $C(u)=\{i\in
L(r):\beta_i^{j(r)}=0\}$. We will make exactly $|C(u)|$ arcs entering into $u$, and give
each such arc a color $i\in C(u)$ (exactly one color from $C(u)$ per arc). Each
arc $(u',u)\in A$, with $u'\in N_{r-1}$ and $i\in C(u)$, will then have the
following meaning: if the constrained sub-problem associated with $u'$ has a
solution, i.e., a coloring $\varphi'$, then we can extend $\varphi'$ into a
solution $\varphi$ to the constrained sub-problem associated with $u$ by giving
color $i$ to vertex $v_r$.

We now give the formal definition of the set $A$. We start with the arcs from
$N_0$ to $N_1$. Let $u=(1,\{\beta_i^j\}_{i=1,\dots,s,j=1,\dots,k})\in N_1$. There is an
arc from $z_i$ (where $i\in L(1)$), to $u$ if and only if $i\in C(u)$; moreover, the color
of its arc is $i$. We now deal with the arcs from $N_{r-1}$ to $N_r$, with $2\le
r \le n$. Let $u=(r,\{\beta_i^j\}_{i=1,\dots,s,j=1,\dots,k})\in N_r$. As we discussed
above, for each $i^*\in C(u)$, there will be an arc from a node $u_{i^*}\in
N_{r-1}$ to $u$, with color $i^*$. Namely,
$u_{i^*}=(r-1,\{\tilde{\beta}_i^j\}_{i=1,\dots,s,j=1,\dots,k})$, where:

\setcounter{equation}{1}
\begin{equation} 
\tilde{\beta}_i^j=
\begin{cases}
\max\{|N(v_r,j)_<|,\beta_i^j\} & i=i^* \\
\max\{0,\beta_i^j-1\} & i\neq i^*, j=j(r) \\
\beta_i^j\ & i\neq i^*, j\neq j(r) \\
\end{cases}
\end{equation}

Note that $u_{i^*}$ is indeed a node of $N_{r-1}$, as the $(sk+1)$-tuple
$(r-1,\{\tilde{\beta}_i^j\}_{i=1,\dots,s,j=1,\dots,k})$ is such that
$0\le\tilde{\beta}_i^j\le \Delta(j)_<$, for each $i,j$ (in fact,
$\beta_i^j\le\Delta(j)_<$, since $u$ is a node of $N_r$).

\begin{lemma}
$G$ admits an $L$ list coloring if and only if $D$ contains a directed path from a
node $z\in N_0$ to $t$. Moreover, if such a path exists, then a list coloring of
$G$ can be obtained by assigning each node $v_r$ $(r \in \{1,\dots,n\})$ the color of the arc of the
path entering into layer $N_r$.
\end{lemma}
\begin{proof}
The proof is analogous to the proof of Lemma 10 in~\cite{bonomo2011bounded} and we omit it here.
\end{proof}

\begin{lemma}
\label{thm:ktcol}
Suppose that for a ($k$-thin) graph $G$ with n vertices we are given an ordering
and a partition of $V(G)$ into $k$ classes that are consistent. Further consider
an instance $(G,L)$ of the list coloring problem. Let $s=\bigl |\bigcup_{v\in
V(G)} L(v)\bigr |$. Then $(G,L)$ can be solved in
$\calO(ns^2k\prod_{j=1,\dots,k}\Delta(j)_<^s)$-time, i.e., 
$\calO(n^{ks+1}s^2k)$-time. 
\end{lemma}
\begin{proof}

By definition, for $r=1,\dots,n-1$, $|N_r|=\prod_{i=1,\dots,k}(\Delta(j)_<+1)^s$. Note
that each node of $D$ has at most $s$ incoming arcs, and each arc can be built
in $\mathcal{O}(sk)$-time. Therefore, $D$ can be built in
$\mathcal{O}(ns^2k\prod_{i=1..k}(\Delta(j)_<+1)^s)$-time. Since $D$ is acyclic,
the reachability problem on $D$ can be solved in linear time. Therefore the list
coloring problem on $G$ can be solved in
$\mathcal{O}(ns^2k\prod_{i=1..k}\Delta(j)_<^s)$-time, that is
$\mathcal{O}(n^{ks+1}s^2k)$-time.
\end{proof}

\begin{lemma}
Let $G$ be a co-comparability graph and $(G,L)$ an instance of the list coloring
problem with the total number of colors $s\ge 2$. Then $(G,L)$ can be solved in
$\mathcal{O}(n^{s^2+1}s^3)$-time, i.e., polynomial time when $s$ is fixed.
\end{lemma}
\begin{proof}
By Lemma~\ref{thm:coco_kthin}, the graph $G$ is $k$-thin. It can be tested in
$\calO(n^3)$ time whether $G$ is
$s$-colorable~\cite{golumbic1977complexity}. If it is $s$-colorable, then by
Lemma~\ref{thm:coco_kthin} we get a comparability ordering and a $k$-partition
of $V(G)$. Moreover, by Lemma~\ref{thm:coco_kthin} we know that $k\le s$. 
Thus, by Lemma~\ref{thm:ktcol}, we can solve the problem in time
$\mathcal{O}(n^3+n^{s^2+1}s^3)=\mathcal{O}(n^{s^2+1}s^3)$.
\end{proof}

\section{Minimum dominating Set}
\label{sec:domset}

In this section, we discuss the minimum dominating set problem on $\graphs{H}$.  The basic idea
behind our algorithms is to reduce the minimum dominating set problem for $H$-graphs to several
minimum dominating set problems on interval graphs, obtained as induced subgraphs of the original
graph.

We start with a useful tool (Lemma~\ref{lem:dominating_set_interval_x}) which states that that one
can compute a dominating set of an interval graph $G$ which is minimum subject to including one or
two of certain special vertices of $G$. This lemma is an essential tool for both of our dominating
set algorithms presented in the subsequent subsections. 

\begin{lemma}\label{lem:dominating_set_interval_x}
Let $G = (V, E)$ be an interval graph and let $C_1, \dots, C_k$ be the left-to-right ordering of the
maximal cliques in an interval representation of $G$. 
\begin{enumerate}
\item
For every $x \in C_1$, a dominating set of $G$ which is minimum subject to including $x$ can be
found in linear time.
\item
For every $x \in C_1$ and $y \in C_k$, a dominating set of $G$ which is minimum subject to including
both $x$ and $y$ can be found in linear time. 
\end{enumerate}
\end{lemma}

\begin{proof}
We provide the proof for the part 1 (the proof of the part 2 follows analogously).  We construct a new graph
$G' = (V', E')$ where $V' = V \cup \{u,u'\}$ and $E' = E \cup \{ux,u'x\}$. Clearly, $G'$ is an
interval graph as certified by the following linear order of its maximal cliques $\{u,x\} = C_0,
C'_0 = \{u',x\}, C_1, \dots, C_k$.  Furthermore, to dominate both $u$ and $u'$ without using $x$, we
would need to include both $u$ and $u'$.  Thus, every minimum dominating set of $G'$ includes $x$,
i.e., we can find such a dominating set in linear time using the standard greedy
algorithm~\cite{agt}. 
\end{proof}

\subsection{Dominating sets in \texorpdfstring{$S_d$}{Sd}-graphs}

Here, we solve the minimum dominating set problem on $\graphs{S_d}$ in $\cFPT$-time, parameterized
by $d$.

\begin{theorem}\label{thm:fpt_star_domination}
For an $S_d$-graph $G$, a minimum dominating set of $G$ can be found in $\calO(dn(n+m)) + 2^d (d +
2^d)^{\calO(1)}$ time when an $S_d$-representation is given. (If such a representation is not given,
we can compute one in $\calO(n^{3.5})$ time by Theorem~\ref{thm:poly_star}.)
\end{theorem}

\begin{proof}
Let $G$ be an $S_d$-graph and let $S'$ be a subdivision of the star $S_d$ such that $G$ has an
$S'$-representation. Let $b$ be the central branching point of $S'$ and let $l_1, \dots, l_d$ be the
leaves of $S'$.  Recall that, by Lemma~\ref{lem:clique_branching_point}, we may assume $b \in
\bigcap\{S'_v : v \in C\}$, for some maximal clique $C$ of $G$.  Let $C_{i,1}, \dots, C_{i,k_i}$ be
the maximal cliques of $G$ as they appear on the branch $P_{(b,l_i]}$, for $i = 1, \dots, d$.

For each $G_i = G[C_{i,1},\dots,C_{i,k_i}]$, we use an interval graph greedy
algorithm~\cite{agt} to find the size $d_i$ of a minimum dominating set in $G_i$.  Let $B_i$ be the
set of vertices of $C$ that can appear in a minimum dominating set of $G_i$.  By
Lemma~\ref{lem:dominating_set_interval_x}, a minimum dominating set $D_i^x$ containing a vertex $x
\in C$ can be found in linear time. Note that $x \in B_i$ if and only if $|D_i^x| = d_i$. Therefore,
every $B_1, \ldots, B_d$ can be found in $\calO(d\cdot n \cdot (n+m))$ time. Let $\calB = \{B_1,
\dots, B_d\}$.

If $B_i$ is empty, then no minimum dominating set of $G_i$ contains a vertex from $C$. So for $G_i$,
we pick an arbitrary minimum dominating set $D_i$.  Note that $D_i$ dominates $C \cap C_{i,1}$
regardless of the choice of $D_i$. Thus, if $\bigcup_{i=1}^d D_i$ dominates $C$, then it is a
minimum dominating set of $G$. Otherwise, $\{x\} \cup \bigcup_{i=1}^d D_i$ is a minimum dominating
set of $G$ where $x$ is an arbitrary vertex of $C$. 

Let us assume now that the $B_i$'s are nonempty (every branch with an empty $B_i$ can be simply
ignored). Let $H$ be a subset of $C$ such that $H \cap B_i$ is not empty, for every $i = 1,
\dots,d$, and $|H|$ is smallest possible. For every branch $P_{(b,b_i]}$, we pick a minimum
dominating set $D_i$ of $G_i$ containing an arbitrary vertex $x_i \in H \cap B_i$. Now, the union
$D_1 \cup \cdots \cup D_d$ is a minimum dominating set of $G$.  It remains to show how to find the
set $H$ in time depending only on $d$.

Finding the set $H$ can be seen as a set cover problem where $\mathcal{B}$ is the ground set.
Namely, we have one set for each vertex $x$ in $C$ where the set of $x$ is simply its subset of
$\mathcal{B}$, and our goal is to cover $\mathcal{B}$. Note, if two vertices cover the same subset
of $\mathcal{B}$ it suffices to keep just one of them for our set cover instance, i.e., giving us at
most $2^d$ sets over a ground set of size $d$. Such a set cover instance can be solved in $2^{d}(d +
2^d)^{\calO(1)}$ time (see Theorem 6.1~\cite{cygan2015parameterized}). 

Thus, we spend $\calO(dn(n+m)) + 2^{d}(d + 2^d)^{\calO(1)}$ time in total. 
\end{proof}

\subsection{Dominating sets in \texorpdfstring{$H$}{H}-graphs}

We turn to $\graphs{H}$, for general fixed $H$. There we solve the problem in $\cXP$-time,
parameterized by $\| H\|$. This latter result can be easily adapted to also obtain $\cXP$-time
algorithms to find a maximum independent set and minimum independent dominating set on $\graphs{H}$
(these algorithms are also parameterized by $\| H\|$); see Corollary~\ref{cor:mis_ids}.

\begin{theorem}\label{thm:xp_domination}
For an $H$-graph $G$ the minimum dominating set problem can be solved in $n^{O(\|H\|)}$ time when an $H$-representation is given as part of the input. 
\end{theorem}

\begin{proof}
Recall that, when $H$ is a cycle, $\graphs{H} = \ca$, i.e., minimum dominating sets can be found
efficiently~\cite{chang1998efficient}.  Thus, we assume $H$ is not a cycle. 

To introduce our main idea, we need some notation.  Consider $G \in \graphs{H}$ and let $H'$ be a
subdivision of $H$ such that $G$ has an $H'$-representation $\{H'_v : v \in V(G)\}$. We distinguish
two important types of nodes in $H'$; namely, $x \in V(H')$ is called \emph{high degree} when it has
at least three neighbors and $x$ is \emph{low degree} otherwise. As usual, the high degree nodes
play a key role.  In particular, if we know the sub-solution which \emph{dominates} the high degree
nodes of $H'$, then the remaining part of the solution must be strictly contained in the low degree
part of $H'$. Moreover, since $H$ is not a cycle, the subgraph $H'_{\leq 2}$ of $H'$ induced by its
low degree nodes is a collection of paths. In particular, the vertices $v$ of $G$ where $H'_v$ only
contains low degree nodes, induce an interval graph $G_{\leq 2}$ and, as such, we can efficiently
find minimum dominating sets on them.  Thus, the general idea here is to first enumerate the
possible sub-solutions on the high degree nodes, then efficiently (and optimally) extend each
sub-solution to a complete solution.  In particular, one can show that in any minimum dominating set
these sub-solutions consist of at most $2\cdot |E(H)|$ vertices (as in
Claim~\ref{lem:mds_in_h-graph} below), and from this property it is not difficult to produce the
claimed $n^{\calO(\|H\|)}$-time algorithm. These ideas are formalized as follows.

We observe that the size of these sub-solutions is ``small''.  Let $D \subseteq V(G)$ be a minimum
dominating set of $G$. For each node $x$ of $H'$, let $V_x = \{v : v \in V(G), x \in H'_v\}$ and
$D_x = \{v : v \in D, x \in H'_v\}$. We further let $D_{\geq 3} = \bigcup \{D_x : \delta_H(x) \geq
3\}$.  We now bound the size of $D_{\geq 3}$ in terms of $H$. 

\begin{claim}\label{lem:mds_in_h-graph}
If $D$ is a minimum dominating set in an $H$-graph $G$, then $|D_{\geq 3}| \leq 2|E(H)|$. 
\end{claim}

\begin{proofclaim}
Consider a high degree node $x$ of $H$ such that $x \in D_{\geq 3}$. For each edge $xx'$ in $H$, let $x = x_1, \ldots, x_k = x'$
be the corresponding path in $H'$. We assign a single vertex $a$ in $D$ to the ordered pair $(x,x')$
such that $H'_a$ contains the longest subpath of $x_1, \ldots, x_k$ including $x = x_1$. Notice that each ordered pair
receives precisely one element of $D$. However, if some element $v$ of $D_{\geq 3}$ was not assigned
to an ordered pair, then it is easy to see that $D$ is not a minimum dominating set (since all
adjacencies achieved by this element are already achieved by the elements we have charged to ordered
pairs).
\end{proofclaim}

By Claim~\ref{lem:mds_in_h-graph}, there are at most $n^{2\cdot|E(H)|}$ possible sets $D_{\geq 3}$.
We now fix one such $D_{\geq 3}$ and describe how to compute a minimum dominating set of $G$
containing it.  Notice that, there can be some difficult decisions we might need to make in this
process. In particular, suppose there is a high degree node $x$ of $H'$ where no vertex from $V_x$
is in $D_{\geq 3}$. It is not clear how we might be able to efficiently choose from ``nearby'' $x$
to dominate these vertices. To get around this case, we simply enumerate more vertices.
Specifically, for each path $P_{[x,y]} = (x, x_1, \ldots, x_k, y)$ in $H'$ where $x$ and $y$ are
high degree nodes (or where $x$ is high degree and $y$ is a leaf), and the $x_i$'s are low degree,
we will pick a ``first'' and ``last'' vertex among the vertices $v$ of $G$ where $H'_v$ is contained
in the subpath $(x_1, \ldots, x_k)$ of $P_{[x,y]}$. That is, for a given $D_{\geq 3}$ we enumerate
all possible subsets of size $2\cdot|E(H)|$ from among the vertices of $G_{\leq 2}$ to act as the
``first'' and ``last'' vertices of each path $P_{[x,y]}$. Clearly, there are at most
$O(n^{2\cdot|E(H)|})$ such subsets. We fix one such subset $D_{\leq 2}$. 

We now have our candidate sub-solutions $D^* = D_{\geq 3} \cup D_{\leq 2}$. There are just some
simple sanity checks we must make on $D^*$ to test if it is a good candidate to be extended to a
dominating set. First, by the definition of $D_{\geq 3}$, it must already dominate every vertex of $G_{\geq 3}$. Second, if there is
some path $P_{[x,y]}$ where $D_{\leq 2}$ contains fewer than two vertices from $P_{[x,y]}$, then
$D^*$ must already dominate every vertex contained in this path. And finally, for every path
$P_{[x,y]}$, for every $v$ with $H'_v$ contained strictly between $x$ and the ``left-end'' of the
``first'' chosen vertex, then $v$ must be dominated by $D_{\geq 3}$. If one of these conditions is
violated, we discard this candidate $D^*$ and go to the next one. 

Finally, what remains to be dominated consists of a collection of disjoint interval graphs where
possibly some sequence of ``left-most'' and ``right-most'' maximal cliques have already been
dominated by $D^*$. Observe that the partially constructed dominating set will consist of one vertex
which reaches the farthest in from the right and one which does the same from the left. Namely, we
can apply Lemma~\ref{lem:dominating_set_interval_x}, to construct a minimum dominating set for each
such interval graph subject to the inclusion of these two special vertices and as such compute a
minimum dominating set of $G$ which contains our candidate partial dominating set. 

This completes the description of the algorithm. From the discussion, we can see that the algorithm
is correct and that the total running time is dominated by the enumeration of the possible sets
$D^*$ plus some additional polynomial factors.  In particular, the algorithm runs in
$n^{\calO(\|H\|)}$ time. 
\end{proof}

We further remark that the above approach can also be applied to solve the maximum independent set
and minimum independent dominating set problems in $n^{O(\|H\|)}$ time. This approach is successful
since these problem can be solved efficiently on interval graphs.  

\begin{corollary}\label{cor:mis_ids}
For an $H$-graph $G$, the maximum independent set problem and minimum independent dominating set
problem can both be solved in $n^{\calO(\|H\|)}$ time.
\end{corollary}

Finally, as we have stated in Section~\ref{sec:results}, in a recent
manuscript~\cite{FominGR-2017arXiv}, \cW{1}-hardness has been shown for both the minimum dominating
set problem and the maximum independent set problem. Moreover, both of these results concern
parameterization by both $\|H\|$ and the solution size. Thus, this classifies the computational
complexity for both of these problems.  It would be interesting to also have \cW{1}-hardness for the
minimum independent dominating set problem.  Additionally, one could make a more fine-grained
examination of the running time and look for lower bounds via ETH.

\begin{problem}
Is the minimum independent dominating set problem \cW{1}-hard on $H$-graphs (parametrized by $\|H\|$
and the solution size)?
\end{problem}

\begin{problem}
Can we obtain some interesting lower bounds using ETH?
\end{problem}

\section{Finding cliques in \texorpdfstring{$H$}{H}-graphs}
\label{sec:clique}

We discuss computational aspects of the maximum clique problem for $H$-graphs, parametrized by
$\|H\|$. Let $\Delta_2$ be the \emph{double-triangle} (see Fig.~\ref{fig:comp2sub}a). First, we show
that the maximum clique problem is \cAPX-hard for $H$-graphs if $H$ contains $\Delta_2$ as a minor
(Theorem~\ref{thm:double-triangle->Subd2_inside}).  In other words, the maximum clique problem is
para-\cNP-hard when parameterized only by $\|H\|$. As a consequence of our reduction, we also show
that if $\Delta_2 \preceq H$, then $\graphs{H}$ is $\cGI$-complete (the graph isomorphism on
$\graphs{H}$ is as hard as the general graph isomorphism problem).  We then turn to cases where the
clique problem can be solved efficiently.  Namely, we consider two cases: one where we have a
``nice'' representation but $H$ is arbitrary, and the other where we restrict $H$ to be a cactus.

\newcommand{\Subd}[1]{\ensuremath{\hbox{\rm \sffamily SUBD}_#1}}
\newcommand{\compSubd}[1]{\ensuremath{\overline{\Subd{#1}}}}

\subsection{Clique (and isomorphism) hardness results} 
\label{sec:clique_and_iso_hardness}

To obtain our hardness results we show that there are graphs $H$ such that the complement of a
$2$-subdivision of every graph is an $H$-graph. The \emph{$2$-subdivision} $G_2$ of a graph $G$ is
the result of subdividing every edge of $G$ exactly two times. The complement of a graph $G$ is
denoted by $\overline{G}$.  We define
$$\compSubd{2} = \{\overline{G_2} : G \text{ is a graph}\}.$$
In other words, $\compSubd{2}$ is the class of complements of $2$-subdivisions of all graphs.

This seemingly esoteric family of graphs is interesting for two reasons. Firstly, the isomorphism
relation on graphs is closed under $k$-subdivision and complement operations. This implies that $G
\cong H$ if and only if $\overline{G_2} \cong \overline{H_2}$. So, the class $\compSubd{2}$ is
$\cGI$-complete.  Secondly, the clique problem is $\cAPX$-hard on $\compSubd{2}$.  More
specifically, Chleb\'ik and Chleb\'ikov\'a~\cite{ChlebikC07-subdiv-hardness} proved that the maximum
independent set problem is \cAPX-hard on the class of $2k$-subdivisions of 3-regular graphs for any
fixed integer $k \geq 0$; in particular, for 2-subdivisions.  Thus, showing that $\compSubd{2}
\subseteq \graphs{H}$, for a fixed $H$, implies that the maximum clique problem is $\cAPX$-hard on
$\graphs{H}$ and that $\graphs{H}$ is $\cGI$-complete. 

\begin{theorem}
\label{thm:double-triangle->Subd2_inside}
If $\Delta_2 \preceq H$, then the maximum clique problem is $\cAPX$-hard for $H$-graphs and
$\graphs{H}$ is $\cGI$-complete. 
\end{theorem}

\begin{proof}
As already mentioned, we prove the theorem by showing $\compSubd{2} \subseteq \graphs{H}$.  Since
$\Delta_2 \preceq H$, the graph $H$ can be partitioned into three connected subgraphs $H_1$, $H_2$,
$H_3$ such that there are at least two edges connecting $H_i$ and $H_j$, for each $i\neq j$.  For
every graph $G$, we show that the complement of its $2$-subdivision has an $H$-representation.

The construction proceeds similarly to the constructions used by Francis et
al.~\cite{Francis2013-multiple-interval}, and we borrow their convenient notation.  Let $G$ be a
graph with vertex set $\{v_1, \ldots, v_n\}$ and edge set $\{e_1, \ldots, e_m\}$. If $e_k \in E(G)$
and $e_k = v_iv_j$ where $i < j$, we define $l(k) = i$ and $r(k) = j$ (as if $v_i$ and $v_j$ were
respectively the \emph{left} and \emph{right} ends of $e_k$). In the 2-subdivision $G_2$ of $G$, the
edge $e_k$ of $G$ is replaced by the path $(v_{l(k)}, a_k, b_k, v_{r(k)})$; see
Fig.~\ref{fig:comp2sub}a and Fig.~\ref{fig:comp2sub}b. 

\begin{figure}[t]
\centering
\includegraphics{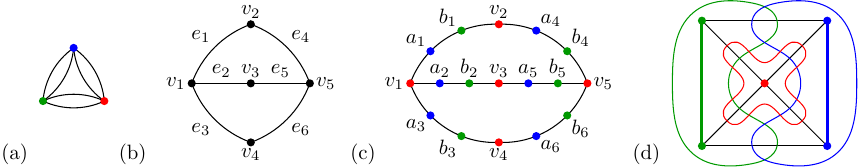}
\caption{
(a) The \emph{double-triangle} graph. (b) A graph $G$. (c) The 2-subdivision $G_2$ of $G$. A
three-clique cover of $\overline{G_2}$ is indicated by colors.  (d) The $4$-wheel graph (which
contains the double-triangle as a minor) and a \emph{sketch} of our $H$-representation of
$\overline{G^*}$. For example, the edges between the green clique and the blue clique are
represented where the green and blue regions intersect.}
\label{fig:comp2sub}
\end{figure}

Note that $\overline{G_2}$ can be covered by three cliques, i.e., $C_v = \{v_1, \ldots, v_n\}$, $C_a
= \{a_1, \ldots, a_m\}$, and $C_b = \{b_1, \ldots, b_m\}$. We now describe a subdivision $H'$ of $H$
which admits an $H$-representation $\{H'_v : v \in V(\overline{G_2})\}$ of $\overline{G_2}$. We
obtain $H'$ by subdividing the six  edges connecting $H_1$, $H_2$, and $H_3$. Specifically: 
\begin{itemize}
\item
We $n$-subdivide the edges connecting $H_1$ to $H_2$ to obtain two paths $P_{12} = (\alpha_0,
\alpha_1, \ldots,$ $\alpha_n, \alpha_{n+1})$, $Q_{12} = (\beta_0, \beta_1, \ldots, \beta_n,
\beta_{n+1})$ where $\alpha_0, \beta_0 \in H_1$ and $\alpha_{n+1}, \beta_{n+1} \in H_2$.
\item
We $n$-subdivide the edges connecting $H_1$ to $H_3$ to obtain two paths $P_{13} = (\gamma_0,
\gamma_1, \ldots,$ $\gamma_n, \gamma_{n+1})$, $Q_{13} = (\eta_0, \eta_1, \ldots, \eta_n,
\eta_{n+1})$ where $\gamma_0, \eta_0 \in H_1$ and $\gamma_{n+1}, \eta_{n+1} \in H_2$. 
\item
We $m$-subdivide the edges connecting $H_2$ and $H_3$ to obtain two paths $P_{23} = (\mu_0, \mu_1,
\ldots,$ $\mu_m, \mu_{m+1})$, $Q_{23} = (\nu_0, \nu_1, \ldots, \nu_m, \nu_{m+1})$ where $\mu_0,
\nu_0, \mu_{m+1}, \eta_{m+1} \in H_2$. 
\end{itemize} 

We now describe each $H_{v_i}$, $H_{a_j}$ and $H_{b_j}$. The idea is that $H'_{v_i}$ will contain
$H_1$ and extend from the ``start'' of $P_{12}$ up to the position $i$, and from the ``start'' of
$Q_{12}$ up to position $(n-i)$.  From the other side, each $H'_{a_j}$ will contain $H_2$ and extend
from the ``end'' of $P_{12}$ down to position $(l(j)+1)$, and from the end of $Q_{12}$ down to
position $(n-l(j)+1)$; an example is sketched in Fig.~\ref{fig:comp2sub}d. In this way, we ensure
that $H'_{a_j}$ does not intersect $H'_{v_{l(j)}}$ while $H'_{a_j}$ does intersect every $H'_{v_i}$
for $i \neq l(j)$. The other pairs proceed similarly, and we describe the subgraphs $H_{v_i},
	H_{a_j}, H_{b_j}$ for each $i \in \{1,\dots,n\}$ and $j \in \{1,\dots,m\}$ as follows: 
\begin{itemize}
\item
$H'_{v_i} = H_1 \cup \{\alpha_1, \ldots, \alpha_i\} \cup \{\beta_1, \ldots, \beta_{n-i}\} \cup
\{\gamma_1, \ldots, \gamma_i\} \cup \{\eta_1, \ldots, \eta_{n-i}\}$.
\item
$H'_{a_j} = H_2 \cup \{\alpha_n, \ldots, \alpha_{l(j)+1}\} \cup \{\beta_n, \ldots,
\beta_{n-l(j)+1}\} \cup \{\mu_1, \ldots, \mu_j\} \cup \{\nu_1, \ldots, \nu_{m-j}\}$. 
\item
$H'_{b_j} = H_3 \cup \{\gamma_n, \ldots, \gamma_{r(j)+1}\} \cup \{\eta_n, \ldots, \eta_{n-r(j)+1}\}
\cup \{\mu_m, \ldots, \mu_{j+1}\} \cup \{\nu_m, \ldots, \nu_{m-j+1}\}$. 
\end{itemize} 
\end{proof}

Some interesting cases remain concerning the maximum clique problem. 
In the next subsection we will prove Theorem~\ref{thm:clique-cactus} which states that, for any cactus $C$,
the clique problem can be solved polynomial time on any $C$-graph.  
Thus, the open cases which remain are when $H$ is not a cactus
(i.e., $H$ contains a diamond as a minor), but $H$ does not satisfy the conditions of
Theorem~\ref{thm:double-triangle->Subd2_inside} (i.e., $H$ does not contain the double-triangle as a minor).  

\begin{problem}
What is the complexity of the maximum clique problem on $H$-graphs in the case when $H$ is not a
cactus and $\Delta_2 \npreceq H$?
\end{problem}

On the other hand, while the isomorphism problem can be solved in linear time on interval graphs and
Helly circular-arc graphs~\cite{curtis2012isomorphism}, split graphs~\cite{lueker1979linear}
are $\cGI$-complete.

\begin{problem}
Let $H$ be a fixed graph such that $\Delta_2 \npreceq H$. What is the complexity of the graph
isomorphism problem on $H$-graphs?
\end{problem}

\subsection{Tractable cases}

Here, we consider two restrictions which allow polynomial-time algorithms for the maximum clique
problem. First, we discuss the case when the $H$-representation satisfies the Helly property. This
is followed by a discussion of the case when $H$ is a cactus. In both situations, we obtain
polynomial-time algorithms. 
 
\heading{Helly H-graphs.}
A \emph{Helly} $H$-graph $G$ has an $H$-representation $\{H'_v : v \in V(G)\}$ such that the
collection $\calH = \{V(H'_v) : v \in V(G)\}$ satisfies the \emph{Helly property}, i.e., for each
sub-collection of $\mathcal{H}$ whose sets pairwise intersect, their common intersection is
non-empty.  Notice that, when $H$ is a tree, every $H$-representation satisfies the Helly property.
When a graph $G$ has a Helly $H$-representation, we obtain the following relationship
between the size of $H$ and the number of maximal cliques in $G$. 

\begin{lemma}\label{lem:helly-h-cliques}
Each Helly $H$-graph $G$ has at most $|V(H)| + |E(H)|\cdot|V(G)|$ maximal cliques.
\end{lemma}
\begin{proof}
Let $H'$ be a subdivision of $H$ such that $G$ has a Helly $H$-representation $\{H'_v : v \in
V(G)\}$. Note that, for each maximal clique $C$ of $G$, $\bigcap_{v \in C} V(H'_v) \neq \emptyset$,
i.e., $C$ corresponds to a node $x_C$ of $H'$.

For every edge $xy \in E(H)$, we consider the
path $P = P_{[x,y]} = (x,x_1,\dots,x_k,y)$ in $H'$. Let $G_{P}$ be the subgraph of $G$
formed by the union of the maximal cliques $C$ of $G$ such that $x_C \in V(P)$.

\begin{claim}
The graph $G_{P}$ is a Helly
circular-arc graph.
\end{claim}

\begin{proof}
Note that if a restriction of $H_v'$, for $v \in V(G_{P})$, to $P$ is disconnected, then it is a disjoint union of two paths containing the end-vertices $x$ and $y$, respectively.
Let $C$ by cycle obtained from $P$ by adding the edge $xy$.
We construct a $C$-representation of $G_P$.
If the restriction of $H_v'$ to $P$ is a subpath of $P$, then we let $C_v$ to be this subpath.
Otherwise, we let $C_v$ to be the restriction of $H_v'$ to $P$ together with the edge $xy$.
Clearly, this is a Helly $C$-representation.
\end{proof}

Now, since Helly circular-arc graphs have at most linearly many maximal
cliques~\cite{gavril1974algorithms}, $G$ has at most $|V(H)| + |E(H)| \cdot |V(G)|$ maximal cliques.
\end{proof}

We can now use Lemma~\ref{lem:helly-h-cliques} to find the largest clique in $G$ in polynomial time.
In fact, we can do this without needing to compute a representation of $G$. In particular, the
maximal cliques of a graph can be enumerated with polynomial delay~\cite{makino2004new}. Thus, since
$G$ has at most linearly many maximal cliques, we can simply list them all in polynomial time and
report the largest, i.e., if the enumeration process produces too many maximal cliques, we know that
$G$ has no Helly $H$-representation. This provides the following theorem. 

\begin{theorem}
\label{thm:helly-h-clique}
The clique problem is solvable in polynomial time on Helly $H$-graphs. 
\end{theorem}

Note that some co-bipartite circular-arc graphs have exponentially many maximal cliques and these
graphs are not contained in Helly $\graphs{H}$, for any fixed $H$.  However, the clique problem is
solvable for circular-arc graphs in polynomial time~\cite{Hsu85-max-clique}. 

\heading{Cactus-graphs.}
The clique problem is efficiently solvable on chordal graphs~\cite{agt} and circular-arc
graphs~\cite{Hsu85-max-clique}. In particular, when $H$ is either a tree or a cycle, the clique
problem can be solved in polynomial-time, independent of $\|H\|$.  In
Theorem~\ref{thm:clique-cactus}, we observe that these results easily generalize to the case when
$G$ is a $C$-graph, for some cactus graph $C$. We define,
$$\cgraphs = \bigcup_{\text{Cactus } C} \graphs{C}.$$

To prove the result we will use the \emph{clique-cutset decomposition}, which is defined as
follows.  A \emph{clique-cutset} of a graph $G$ is a clique $K$ in $G$ such that $G - K$ has
more connected components than $G$. An \emph{atom} is a graph without a clique-cutset.  An
\emph{atom of a graph $G$} is a maximal induced subgraph $A$ of $G$ which is an atom. A \emph{clique-cutset
decomposition} of $G$ is a set $\{A_1, \ldots, A_k\}$ of atoms of $G$ such that $G = \bigcup_{i=1}^k
A_i$ and for every $i,j$, $V(A_i) \cap V(A_j)$ is either empty, or induces a clique in $G$.
Algorithmic aspects of clique-cutset decompositions were studied by Whitesides~\cite{Whi1984} and
Tarjan~\cite{Tar1985}. In particular, if $k\leq n$, then for any graph $G$ a clique-cutset
decomposition $\{A_1, \ldots, A_k\}$ of $G$ can be computed in $O(n^2+nm)$~\cite{Tar1985}.
Additionally, to solve the clique problem on a graph $G$ it suffices to solve it for each atom of
$G$ from a clique-cutset decomposition~\cite{Whi1984,Tar1985}.  Theorem~\ref{thm:clique-cactus} now
follows from the following easy lemma and the fact that the clique problem can be solved in
polynomial time for circular-arc graphs~\cite{Hsu85-max-clique}. 

\begin{lemma}
\label{obs:cactus-atoms}
Let $C$ be cactus and let $G \in \graphs{C}$. Then each atom $A$ of $G$ is a circular-arc graph.
\end{lemma}

\begin{proof}
Consider an $C$-representation $\{C_v' : v \in V(G)\}$ of $G$.  Now, let $C|_A = \bigcup_{v \in
V(A)} C_v'$. Clearly, if $C|_A$ is a path or a cycle, then we are done. Otherwise, $C|_A$ must
contain a cut-node $x$.  Let $X_1, \ldots, X_t$ be the components of $H|_A - \{x\}$, and let
$S$ be the vertices of $A$ whose representations contain $x$.  Note that $S$ is a clique in $A$.
Moreover, since $A$ is an atom, $S$ is not a clique-cutset.  Thus, there is a component $X_j$ such
that the subgraph $C^*$ of $C$ induced by $V(X_j) \cup \{x\}$ provides a representation of $A$. In
particular, if $C^*$ is either a cycle, or a path we are again done.  Moreover, when $C^*$ is
neither a path, nor a cycle, repeating this argument on $C^*$ provides a smaller subgraph of $C$, on
which $A$ can be represented, i.e., this eventually produces either a path, or cycle.
\end{proof}

\begin{theorem}
\label{thm:clique-cactus}
The clique problem can be solved in polynomial time on $\cgraphs$. 
\end{theorem}

\section{FPT results via clique-treewidth graph classes}
\label{sec:treewidth}

The concept of \emph{treewidth} was introduced by Robertson and Seymour~\cite{robertson1984graph}.
A {\em tree decomposition} of a graph $G$ is a pair $({X}, T)$, where $T$ is a tree and ${X}=\{{X}_i
\mid i\in V(T) \}$ is a family of subsets of $V(G)$, called \emph{bags}, such that (1) for all $v
\in V(G)$, the set of nodes $T_v = \{i \in V(T) \mid v \in {X}_i\}$ induces a non-empty connected
subtree of $T$, and (2) for each edge $uv \in E(G)$ there exists $i \in V(T)$ such that both $u$ and
$v$ are in ${X}_i$.  The maximum of $|{X}_i|-1$, $i\in V(T)$, is called the {\em width} of the tree
decomposition.  The {\em treewidth}, $\tw(G)$, of a graph $G$ is the minimum width over all tree
decompositions of $G$.

An easy lower bound on the treewidth of a graph $G$ is the size of the largest clique in $G$, i.e.,
its clique number $\omega(G)$. 
This follows from the fact that each edge of $G$ belongs to some bag
of $T$ and that a collection of pairwise intersecting subtrees of a tree must have a common
intersection (i.e., they satisfy the Helly property). 
With this in mind, we say that a graph class
$\calG$ has the \emph{clique-treewidth property}\footnote{In our prior work~\cite{ChaplickZ17}, we referred to this as being \emph{treewidth-bounded}, but have changed the name to be consistent with other parameter-treewidth bounds given in \emph{bidimensionality theory}~\cite{parametertreewidth2004}.} if there is a function $f: \mathbb{N}\rightarrow
\mathbb{N}$ such that for every $G \in \calG$, $\tw(G) \leq f(\omega(G))$.  This concept
generalizes the idea of $\calG$ being \emph{$\chi$-bounded}, namely, that the \emph{chromatic
number} $\chi(G)$ of every graph $G \in \calG$ is bounded by a function of the clique number
of $G$. In particular, the chromatic number of a graph $G$ is bounded by its treewidth since a tree
decomposition $(X,T)$ of $G$ is a $T$-representation  of a chordal supergraph $G'$ of $G$ where $\omega(G') =
\tw(G)+1$, i.e., $\chi(G') = \tw(G)+1$ since chordal graphs are perfect.  
It was recently shown that the graphs which do not contain \emph{even holes} (i.e., cycles of length $2k$ for any $k\geq 2$)
and \emph{pans} (i.e., cycles with a single pendent vertex attached) as induced subgraphs have their
treewidth bounded by $f(\omega) = 3\omega/2 - 1$~\cite{cameron2015structure}. 
For a function $f: \mathbb{N} \rightarrow \mathbb{N}$, we use $\calG_f$ to denote the class of
graphs $G$ where $\tw(G) \leq f(\omega(G))$.  Each class \graphs{H} is known to be a subclass of
$\calG_f$ for certain linear functions $f$, as in the following lemma.

\begin{lemma}[Bir\'{o}, Hujter, and Tuza~\cite{biro1992precoloring}]
\label{lem:bounded_tw}
For every $G \in \graphs{H}$, $\tw(G) \leq (\tw(H)+1) \cdot \omega(G) -1$, i.e., $\graphs{H}$ is a
subclass of $\calG_{f_H}$, where $f_H(\omega) = (\tw(H)+1) \cdot \omega -1$. 
\end{lemma}

We leverage any clique-treewidth-property (e.g., as in Lemma~\ref{lem:bounded_tw}) together with some existing algorithms to to classify the
$k$-coloring and $k$-clique problems as \cFPT\ on the $\calG_{f}$ classes (e.g., on $\graphs{H}$ classes as well). 
We first consider the $k$-clique problem.

\begin{theorem}\label{thm:fpt-k-clique}
For any monotone computable function $f: \mathbb{N} \rightarrow \mathbb{N}$, the $k$-clique problem can be
solved in $2^{\calO(f(k))}\cdot n$ time for $G \in \calG_f$. Thus, for
$\graphs{H}$, the $k$-clique problem can be solved in $2^{\calO(\tw(H)\cdot k)}\cdot n$ time.   
\end{theorem}

\begin{proof}
To test if $G$ contains a $k$-clique, we first try to generate a tree decomposition of $G$ with
width roughly $f(k)$ via a recent algorithm~\cite{DBLP:journals/siamcomp/BodlaenderDDFLP16}, which,
for any given graph $G$ and number $t$, provides a tree decomposition of width at most $5\cdot t$ or
states that the treewidth of $G$ is larger than $t$. This algorithm runs in $2^{\calO(t)}\cdot
n$ time. 
If this algorithm does not produce a tree decomposition, then $G$ must contain a $k$-clique, and we are done. 
Otherwise, we obtain a tree decomposition $({X}, T)$ of $G$ of width $5\cdot f(k)$. 
Note that, an easy property of tree decompositions is that, for every clique $K$, there is a bag
which contains the vertices of $K$. 
In particular, to check if $G$ has a $k$-clique it suffices to check whether each the subgraph induced by a bag of $G$ contains a $k$-clique. 
This can obviously be done in $2^{\calO(f(k))} \cdot n$ time by brute-force. 
Thus, we have $2^{\calO(f(k))}\cdot n$ time in total as needed.
\end{proof}

For each fixed $k \geq 3$, it is known that testing \emph{$(k,k)$-pre-colouring extension} (see
Section~\ref{sec:h-graphs} for a definition) for $G \in \graphs{H}$ can be done in \cXP~time
\cite{biro1992precoloring}.  The authors combine Lemma~\ref{lem:bounded_tw} together with a simple
argument to obtain the result.  We use a similar argument together with a more recent result
regarding bounded treewidth graphs to observe that an even more general problem, \emph{list
$k$-coloring} (where each list is a subset of $\{1, \ldots, k\}$), is \cFPT\ on graph class
satisfying the clique-treewidth property, 
and therefore, also on \graphs{H}.

\begin{theorem}\label{thm:fpt-k-list-col}
For any monotone computable function $f: \mathbb{N} \rightarrow \mathbb{N}$, the list-$k$-coloring problem can be solved
in $k^{\calO(f(k))}\cdot n$ time for $G \in \calG_f$. Thus,
for $\graphs{H}$, the list-$k$-coloring problem can be solved in $k^{\calO(\tw(H)\cdot k)}\cdot n$ time.   
\end{theorem}

\begin{proof}
For fixed $k$, clearly, if $G$ contains a clique of size $k+1$ then $G$ has no $k$-coloring, i.e.,
no list-$k$-coloring, regardless of the lists.  We use Theorem~\ref{thm:fpt-k-clique} to test for such
a clique, and reject if one is found.  Otherwise, we have a $5\cdot f(k)$-width tree decomposition,
and this time we use it to solve the list-$k$-coloring problem via the known $\calO(k^{t+2} \cdot n)$-time
algorithm when given a width $t$ tree decomposition~\cite{JansenScheffler-k-list-col-1997}. 
Thus, list-$k$-coloring can be solved in $(2^{\calO(f(k))}+k^{\calO(f(k))})\cdot n$ time on $\calG_f$. 
\end{proof}

Some further natural open questions remain regarding these results. For example, what other problems
can be approached on graph classes with the clique-treewidth property?  Can we obtain
polynomial-size kernels for the $k$-clique or list-$k$-coloring problems on $\graphs{H}$ or more
generally on graph classes with the clique-treewidth property? 
The kernelization question has already been partially answered for the $k$-clique problem. 
Namely, on $H$-graphs, it was recently shown~\cite{FominGR-2017arXiv} that the $k$-clique 
admits a polynomial kernel in terms of $\|H\|$ and $k$, but the kernelization requires an $H$-representation 
to be given as part of the input. In contrast, our $\cFPT$ algorithm for $k$-clique (while also 
parameterized by both $\|H\|$ and $k$) does not need an $H$-representation.

\begin{problem}
Can the kernelization for $k$-clique be done without an $H$-representation a part of the input?
\end{problem}

\begin{problem}
Can we obtain polynomial-size kernel for the list-$k$-coloring problem on $H$-graphs?
\end{problem}

\section{Minimal separators}
\label{sec:min_sep}

For a connected graph $G$, a subset $S$ of $V(G)$ is a \emph{minimal separator} when $G$ has
vertices $u$ and $v$ belonging to distinct components of $G - S$ such that no proper subset of $S$
disconnects $u$ and $v$ -- here, we say that $S$ is \emph{minimal $(u,v)$-separator}.  We denote the
set of all minimal separators in $G$ by $\calS(G)$.  Minimal separators are a commonly studied aspect of
many graph classes~\cite{BouchitteT01,GaspersM15,agt,KloksKW98}. Two particularly relevant cases
include the fact that chordal graphs have at most $n$ minimal separators~\cite{agt}, and that
circular-arc graphs have at most $2n^2-3n$ minimal separators~\cite{KloksKW98}.

Recently, several algorithmic results have been developed, where the runtime depends on the number
of minimal separators in the input graph. The main result in this direction is the one by Fomin,
Todinca, and Villanger~\cite{FominTV15}, which is phrased in terms of \emph{potential maximal
cliques}, but can also be phrased in terms of minimal separators since the number of potential
maximal cliques in a graph $G$ is bounded by $n |\calS(G)|^2$ (see Proposition~2.8
in~\cite{FominTV15}). Roughly, in~\cite{FominTV15} the authors show that a large class of problems
can be solved in time polynomial in the number of minimal separators of the input graph. These
problems include several standard combinatorial optimization problems, e.g., maximum independent set
and \emph{maximum induced forest}\footnote{i.e., minimum feedback vertex set}.

The class of problems considered in~\cite{FominTV15} is formalized as follows.  Consider a fixed
integer $t\geq 0$, and a formula $\varphi$ expressed in \emph{counting extended monadic second order
logic} (CMSO)\footnote{Informally, CMSO consists of all logic formulas with quantifiers over
vertices, edges, edge sets and vertex sets, and counting modulo constants. For more information on
this logic see, e.g.,~\cite{Courcelle:2012}. Note: in~\cite{Courcelle:2012}, this logic is
abbreviated by CMS$_2$ instead of CMSO as in~\cite{FominTV15}.}. For an input graph $G$, the goal is
to find a maximum size subset $X \subseteq V(G)$ satisfying: there is $F \subseteq V(G)$ such that
$X \subseteq F$, the subgraph $G[F]$ has treewidth at most $t$, and the structure $(G[F], X)$ models
$\varphi$. The graph $G[X]$ is called \emph{maximum induced subgraph of treewidth $\leq t$
satisfying $\varphi$}. The main result of~\cite{FominTV15} is that this problem can be solved in
time $O(|\calS|^2 n^{t+5} f(t,|\varphi|))$ where $f$ is a computable function.

Now, we prove that each $H$-graph has $n^{O(\|H\|)}$ minimal separators; see
Theorem~\ref{thm:min_sep}. We obtain Corollary~\ref{cor:min_sep_cms} by applying the
meta-algorithmic result of Fomin, Todinca, and Villanger. Subsequently, we consider the case of
$H$-graphs when $H$ is a cactus and observe a much smaller bound on the number of minimal
separators, in particular, $O(\|H\|n^2)$; see Theorem~\ref{thm:min_sep_cactus}. Similarly, by
applying the meta-algorithmic result we obtain Corollary~\ref{cor:min_sep_cactus_cms}: for
cactus-graphs, the maximum induced subgraph of treewidth $t$ modelling $\varphi$ can be solved in
polynomial time.

\begin{theorem}\label{thm:min_sep}
Let $G$ be a connected $H$-graph. Then $G$ has $n^{O(|E(H)|)}$ minimal separators.
\footnote{A similar result with a slightly better bound is given in a recent manuscript,
see~\cite{FominGR-2017arXiv}. Our proof and theirs seem to follow similar reasoning, but have been
obtained independently, as also noted in~\cite{FominGR-2017arXiv}.}
\end{theorem}

\begin{proof}
We show that each minimal separator arises from vertices of $G$ such that their
representations contain a small number of edges of the subdivision $H'$. Then we count all such
subsets of edges of $H'$. 

Let $H'$ be a subdivision of $H$ certifying that $G$ is an $H$-graph.  Let $H^*$ be the subgraph of
$H'$ formed by the union of the representations of the vertices of $G$, i.e.,
$$H^* = \bigcup_{x \in V(G)} H'_x.$$
Observe that, since $G$ is connected, $H^*$ must be also connected. Moreover, for any minimal
$(u,v)$-separator $S$, the graph $H_S^* = \bigcup_{x \in V(G) \setminus S} H'_x$ is not connected.
Now, since $S$ is an $(u,v)$-separator, there are distinct components $Z_u^*$ and $Z_v^*$ of $H_S^*$
such that $H'_u$ is a subgraph of $Z_u^*$ and $H'_v$ is a subgraph of $Z_v^*$.

Observe that, since $S$ is minimal, then if $x \in S$, then the representation $H_x'$ contains an edge $ab$ of $H^*$ such that either $a \in V(Z_u^*)$ and $b\notin V(Z_u^*)$, or $a \in V(Z_v^*)$ and $b\notin V(Z_v^*)$.
Namely, there is a set $E_S$ of edges of $H^*$ such that $S$ is precisely the set of vertices $x$ of
$G$ where $H'_x$ contains an edge of $E_S$. Moreover, for each edge of $H$, at most two edges from
its path in $H'$ occur in $E_S$.

To bound the number of all possible minimal separators in $G$, it
suffices to enumerate all possible subsets $E$ of $E(H')$ where, for each edge of $H$, we pick at
most two edges from its path in $H'$. Here, the candidate separator $S$ would simply be all vertices $x$
of $G$ for which $H'_x$ contains an edge of $E$.  Thus, since each edge of $H$ will be subdivided at
most $2n-1$ times (since $2n$ nodes are sufficient to accommodate any circular-arc representation),
we obtain that the number of minimal separators in $G$ is at most $$\left (\binom{2n}{2} + \binom{2n}{1}
+\binom{2n}{0}\right )^{|E(H)|} = n^{\calO(|E(H)|)}.$$
\end{proof}

\begin{corollary}\label{cor:min_sep_cms}
Let $H$ be a fixed graph. For every $H$-graph $G$, $t \geq 0$, and every CMSO formula
$\varphi$, a maximum induced subgraph of treewidth $\leq t$ modelling $\varphi$ can be found in time
$O(n^{c|E(H)|} n^{t+5} f(t,\varphi))$, where $c$ is a constant and $f$ is a computable function.
\end{corollary}

\begin{theorem}\label{thm:min_sep_cactus}
Let $G$ be a connected $C$-graph, where $C$ is a cactus graph. Then $G$ has at most $|E(C)|(2n^2+n)$
minimal separators. 
\end{theorem}

\begin{proof}
The reasoning here follows similarly to the proof of Theorem~\ref{thm:min_sep}. Namely, if we
consider a minimal $(u,v)$-separator, we again find the components $Z_u^*$ and $Z_v^*$ in the
subdivision $H'$. However, since $H'$ is a cactus, we can now look more closely at the edges which
are incident to $Z_u^*$ and $Z_v^*$ but contained in neither. In particular, it is easy to see that
among all such edges incident to $Z_u^*$, there are at most two edges $e_1,e_2$ which are actually
important to ensure that there is no path from $H'_u$ from $H'_v$. In other words, our set $E_S$
consists of at most two edges of $H'$. Moreover, these two edges must belong to the same cycle of
$H'$. Finally, since each cycle of $H'$ forms a circular-arc graph, it never needs to contain more
than $2n$ nodes, i.e., also $2n$ edges. Thus, since $H$ contains at most $|E(H)|$ cycles, the number
of minimal separators in $G$ is at most
$|E(H)|\left (\binom{2n}{2}+\binom{2n}{1}\right ) \leq |E(H)|(2n^2+n)$.
\end{proof}

\begin{corollary}\label{cor:min_sep_cactus_cms}
Let $C$ be a cactus. For every $G \in \graphs{C}$, $t \geq 0$ and every CMSO formula $\varphi$, a
maximum induced subgraph of treewidth $\leq t$ modelling $\varphi$ can be found in time $O(|E(C)|^2
n^{t+9} f(t,\varphi))$, where $f$ is a computable function.
\end{corollary}

As we have mentioned, two recent manuscripts~\cite{FominGR-2017arXiv,JaffkeKT-2017arXiv} have
obtained \cW{1}-hardness results for both the maximum independent set problem and the minimum
feedback vertex set problem (respectively) when parameterized by $\|H\|$ and the solution size. In
both results, the graphs $H$ which are used have progressively larger clique minors. These indicate
that the \cXP-time results of Corollary~\ref{cor:min_sep_cms} are extremely unlikely to be improved
to $\cFPT$-time, even when adding the solution size as an additional parameter.  On the other hand, as in
Corollary~\ref{cor:min_sep_cactus_cms}, when $H$ is a cactus (i.e., diamond-minor free), these
problems (and many more) can be solved in polynomial time in both $\|H\|$ and the size of the input
graph.

\begin{problem}
For which classes $\mathcal{H}$ (besides the cacti), can one similarly bound the number of minimal
separators by a polynomial in terms of $\|H\|$ and $\|G\|$ where $H \in \mathcal{H}$ and $G$ is an
$H$-graph?
\end{problem}

\section*{Acknowledgements}

We would like to thank Pavel Klav\'{i}k for suggesting to study of $H$-graphs and for several helpfull
discussions. We would also like to thank the DIMACS REU 2015 program, held at the Rutgers
University, where the whole project started.

\bibliographystyle{alpha}
\bibliography{h-graphs}

\end{document}